\documentclass[aps,prd,
superscriptaddress,
showkeys,
showpacs]{revtex4}

\usepackage[english]{babel}
\usepackage{amsmath,bm}
\usepackage{amssymb}
\usepackage{enumitem}
\usepackage{textcomp}
\usepackage{graphicx}
\usepackage{dsfont}
\usepackage{color}
\usepackage{hyperref}
\usepackage{mathrsfs}
\usepackage[normalem]{ulem}
\begin{document}

\title{Schwinger production of scalar particles during and after inflation \\ from the first principles}

\author{O.O.~Sobol}
\email{oleksandr.sobol@epfl.ch}
\affiliation{Institute of Physics, Laboratory for Particle Physics and Cosmology (LPPC), \'{E}cole Polytechnique F\'{e}d\'{e}rale de Lausanne (EPFL), CH-1015 Lausanne, Switzerland}
\affiliation{Physics Faculty, Taras Shevchenko National University of Kyiv, 64/13, Volodymyrska Str., 01601 Kyiv, Ukraine}

\author{E.V.~Gorbar}
\affiliation{Physics Faculty, Taras Shevchenko National University of Kyiv, 64/13, Volodymyrska Str., 01601 Kyiv, Ukraine}
\affiliation{Bogolyubov Institute for Theoretical Physics, 14-b, Metrologichna Str., 03680 Kyiv, Ukraine}

\author{A.I.~Momot}
\affiliation{Physics Faculty, Taras Shevchenko National University of Kyiv, 64/13, Volodymyrska Str., 01601 Kyiv, Ukraine} 

\author{S.I.~Vilchinskii}
\affiliation{Institute of Physics, Laboratory for Particle Physics and Cosmology (LPPC), \'{E}cole Polytechnique F\'{e}d\'{e}rale de Lausanne (EPFL), CH-1015 Lausanne, Switzerland}
\affiliation{D\'{e}partement de Physique Th\'{e}orique, Center for Astroparticle Physics, Universit\'{e} de Gen\`{e}ve, 1211 Gen\`{e}ve 4, Switzerland}
\affiliation{Physics Faculty, Taras Shevchenko National University of Kyiv, 64/13, Volodymyrska Str., 01601 Kyiv, Ukraine}

\date{\today}
\pacs{04.62.+v, 47.75.+f, 98.80.Cq, 98.62.En}

\begin{abstract}
By using the first-principles approach, we derive a system of three quantum kinetic equations governing the production and evolution of charged scalar particles by an electric field in an expanding universe.
Analyzing the ultraviolet asymptotic behavior of the kinetic functions, we found the divergent parts of the electric current and the energy-momentum tensor of the produced particles and determined the corresponding counterterms.
The renormalized system of equations is used to study the generation of electromagnetic fields during and after inflation in the kinetic coupling model
$\mathcal{L}_{\rm EM}=-(1/4)f^{2}(\phi)F_{\mu\nu}F^{\mu\nu}$ with the Ratra coupling function $f=\exp(\beta\phi/M_{p})$. It is found that the electric current of created particles is retarded with 
respect to the electric field. This leads to an oscillatory behavior of both quantities in agreement with the results obtained previously in phenomenological kinetic and hydrodynamical approaches.
\end{abstract}

\keywords{Schwinger effect, charged scalar particles, quantum kinetic equation}

\maketitle

\section{Introduction}
\label{sec-intro}

Observations of the gamma rays from distant blazars demonstrate the presence of magnetic fields in cosmic voids \cite{Tavecchio:2010,Ando:2010,Neronov:2010,Dolag:2010,Dermer:2011,Taylor:2011,Caprini:2015}. 
Although a number of generation mechanisms has been proposed in the literature (see Refs.~\cite{Kronberg:1994,Grasso:2001,Widrow:2002,Giovannini:2004,Kandus:2011,Durrer:2013,Subramanian:2016} for review), an 
extremely large coherence length of these fields strongly favors the inflationary magnetogenesis \cite{Turner:1988,Ratra:1992,Garretson:1992,Dolgov:1993} as their most natural origin. During inflation the coupling 
of an electromagnetic field to an evolving inflaton field or curvature scalar breaks the conformal invariance of the Maxwell action and generates magnetic as well as electric fields. Since the produced electric fields are 
typically larger or at least of the same magnitude as the magnetic ones, the role and impact of these fields on the inflationary magnetogenesis is very important.

The first immediate consequence of the existence of strong electric fields is the formation of plasma of charged particles and antiparticles due to the Schwinger effect \cite{Sauter:1931,Heisenberg:1936,Schwinger:1951}. 
Indeed, it is well known that vacuum becomes unstable with respect to the creation of pairs of charged particles in a strong electric field $\mathbf{E}$ with the production rate in flat spacetime $\propto E^2\exp(-\pi m^2c^3/|eE|\hbar)$
\cite{Nikishov:1970,Cohen:2008}. Obviously, the pair production is exponentially suppressed unless the electric field exceeds the critical value, which is $E_{\rm cr}=m^2_ec^3/(e\hbar) \simeq 1.3\times 10^{16}\,{\rm V/cm}$ 
for the lightest charged particle in the Standard Model, which is the electron. Since the critical field value is extremely large, the Schwinger pair production was never observed experimentally.

Created particles are rapidly accelerated by the electric field and screen it. Therefore, when the Schwinger process becomes efficient, the backreaction of created 
particles on the electric field cannot be neglected. This means that a usual study of the Schwinger effect in adiabatically (compared to the universe expansion rate) evolving and homogeneous electric field in de 
Sitter spacetime
\cite{Kobayashi:2014, Froeb:2014, Bavarsad:2016, Stahl:2016a, Stahl:2016b, Hayashinaka:2016a,  Hayashinaka:2016b,Sharma:2017,Bavarsad:2018,Geng:2018,Hayashinaka:2018, Hayashinaka:thesis,Giovannini:2018a,Banyeres:2018,Domcke:2018,Domcke:2019} 
is not sufficient. Nevertheless, such a study admits an exact analytic solution which was used to constrain different models of magnetogenesis
\cite{Kobayashi:2014,Banyeres:2018,Domcke:2018,Domcke:2019,Stahl:2018,  Kitamoto:2018,Sobol:2018,Kobayashi:2019,Sobol:2019,Shtanov:2020}, to describe reheating \cite{Tangarife:2017} or to quantify a qualitative impact of the Schwinger 
effect on primordial spectra \cite{Chua:2019} (see also Ref.~\cite{Shakeri:2019} for the case of complex inflaton field).

In our recent work \cite{Gorbar:2019}, we attempted a phenomenological approach to the kinetic analysis of the Schwinger effect in an expanding universe and used a model expression for the crucially important Schwinger 
source term in the Boltzmann equation which describes the production of charged particles. Such a source term was previously utilized in the description of the Schwinger effect in heavy-ion collisions and laser 
beams \cite{Kajantie:1985,Gatoff:1987,Kluger:1991,Kluger:1992,Rau:1994,Schmidt:1998,Kluger:1998,Bloch:1999,Schmidt:1999,Alkofer:2001,Blaschke:2019}. It reproduces the correct pair creation rate per unit time and takes into account the stimulated pair production for bosons and the Pauli blocking for fermions. 
In our study \cite{Gorbar:2019}, we found that due to the inertia of charge carriers the electric current remains nonzero even when the electric field crosses zero and changes the sign. This means that the current is retarded with respect to the field and cannot be expressed in the Ohmic form $j=\sigma E$ which is local in time. As a result, both quantities oscillate 
in time with the characteristic frequency comparable or even larger than the expansion rate.

The strongest drawback of our analysis in Ref.\cite{Gorbar:2019} was the use of a phenomenological Schwinger source term adapted from the studies in flat spacetime \cite{Kajantie:1985,Gatoff:1987,Kluger:1991,Kluger:1992}. In this paper, we go beyond this
approximation and apply the first-principles quantum kinetic approach \cite{Rau:1994,Schmidt:1998,Kluger:1998,Bloch:1999,Schmidt:1999,Alkofer:2001,Blaschke:2019} in order to calculate the Schwinger source for a time-dependent
electric field in an expanding universe. This provides the main motivation for the present study. Our quantum kinetic calculations encounter the usual ultraviolet (UV) divergences in the electric current and the energy-momentum 
tensor. Performing the renormalization procedure, we obtain a coupled system of equations governing the joint evolution of the inflaton, electromagnetic field, and produced charged particles. This is the
central result of our study.

As a test bed for our approach, we consider a simple model of the kinetic coupling of the electromagnetic field to the inflaton field via the term $-(1/4)f^2(\phi)F_{\mu\nu}F^{\mu\nu}$ introduced 
by Ratra \cite{Ratra:1992} and revisited many times in the
literature \cite{Giovannini:2001,Bamba:2004,Martin:2008,Demozzi:2009,Kanno:2009,Ferreira:2013,Ferreira:2014,Vilchinskii:2017,Sharma:2017b,Shtanov:2018}. The coupling function is taken in the Ratra form $f=\exp(\beta\phi/M_{p})$ which is a decreasing function of time during inflation. It does not cause a strong-coupling problem \cite{Demozzi:2009} and leads to the production of strong electric fields \cite{Martin:2008,Demozzi:2009,Kanno:2009,Sobol:2018}, much stronger than the magnetic ones. In order to obtain specific results, the numerical analysis is performed in the Starobinsky model of inflation \cite{Starobinsky:1980}. It belongs to the class of plateau models strongly favored by the cosmic microwave background (CMB) observations \cite{Planck:2018inf}.  The kinetic approach and the results obtained in this paper can be easily generalized to other inflationary models.

This paper is organized as follows. A system of quantum Vlasov equations describing the Schwinger pair production of scalar particles is derived in Sec.~\ref{sec-kinetic}. The UV asymptotic behavior 
of the kinetic functions is studied and counterterms are determined in Sec.~\ref{sec-UV}. The application of the renormalized system of equations to the generation of electromagnetic fields during and after inflation is 
considered in Sec.~\ref{sec-kinetic-coupling}. Our numerical results are presented in Sec.~\ref{sec-numerical} and conclusions are given in Sec.~\ref{sec-concl}. The UV asymptotics 
of the kinetic functions are calculated in Appendix~\ref{app-UV}. The dimensional regularization and subtraction scheme are discussed in Appendix~\ref{app-dimensional}. Some details of the calculation of the
energy-momentum tensor for scalar charged particles are given in Appendix~\ref{app-EMT}. Throughout the work we use the natural units and set $\hbar=c=1$.

\section{Quantum kinetic description of Schwinger pair production}
\label{sec-kinetic}

The Schwinger pair production in the early universe can occur if the two main ingredients are present: (i) a strong electric field and (ii) at least one quantum charged field. In our work we will focus on the case
where the electric field is produced in an inflationary magnetogenesis scenario and the charged field is a scalar field nonminimally coupled to gravity (the extension to the case of a charged fermion field 
or multiple scalar and fermion fields is rather straightforward and will be considered elsewhere). The corresponding action reads
\begin{equation}
	\label{action}
	S=\int d^{4}x\sqrt{-g}\left[-\frac{M_{p}^{2}}{2}R+ \mathcal{L}_{\rm inf} +\mathcal{L}_{\rm EM}+\mathcal{L}_{\rm ch}\right],
\end{equation}
where $g={\rm det}(g_{\mu\nu})$, $g_{\mu\nu}={\rm diag\,}\{1,-a^2,-a^2,-a^2\}$ is the metric of a spatially flat Friedmann-Lema{\^{i}}tre-Robertson-Walker (FLRW) universe in the cosmic time, $a=a(t)$ is the scale factor, $M_{p}=(8\pi G)^{-1/2}=2.43\times 10^{18}\,{\rm GeV}$ is the reduced Planck mass, and $R$ is the curvature scalar. The inflationary part consists of a single scalar inflaton field with potential $V(\phi)$,
\begin{equation}
\mathcal{L}_{\rm inf}=\frac{1}{2}g^{\mu\nu}\partial_{\mu}\phi\partial_{\nu}\phi -V(\phi).
\end{equation}
The electromagnetic sector
\begin{equation}
\label{lagr-EM-gen}
\mathcal{L}_{\rm EM}=-\frac{1}{4}F_{\mu\nu}F^{\mu\nu}+\mathcal{L}_{\rm int}(A_{\mu},\phi)
\end{equation}
contains the Maxwell term as well as the term $\mathcal{L}_{\rm int}$ describing the interaction of the electromagnetic field with the inflaton leading to the generation of electromagnetic fields (in our 
numerical analysis, we use the kinetic coupling model presented in Sec.~\ref{sec-kinetic-coupling}). Finally, the charged scalar field is described by the Lagrangian density
\begin{equation}
\label{lagr-chi}
\mathcal{L}_{\rm ch}=g^{\mu\nu}(\mathcal{D}_{\mu}\chi)^{\dagger}(\mathcal{D}_{\nu}\chi)-(m^{2}-\xi R)|\chi|^{2},
\end{equation}
where $\mathcal{D}_{\mu}=\partial_{\mu}-ie A_{\mu}$ is the covariant derivative acting on the scalar field with charge $e$ and mass $m$. Coupling constant $\xi$ quantifies the nonminimal coupling of the scalar field $\chi$ to gravity, and $|\chi|^{2}=\chi^{\dagger}\chi$.

In this and the next sections, we derive a system of quantum Vlasov equations governing the production and evolution of charged scalar particles by an electric field in an expanding universe without specification of a model of magnetogenesis. Since the charged scalar field interacts only with the electromagnetic field and the curvature scalar, it suffices for our aims to assume that the time evolution of both of them is 
given. Then varying action (\ref{action}) with respect to $\chi^{\dagger}$, we obtain the equation of motion for the scalar field
\begin{equation}
\label{eq-scalar}
\frac{1}{\sqrt{-g}}\mathcal{D}_{\mu}\left[\sqrt{-g}g^{\mu\nu}\mathcal{D}_{\nu}\chi\right]+(m^{2}-\xi R)\chi=0
\end{equation}
or in more explicit form
\begin{equation}
\label{eq-scalar-2}
\ddot{\chi}+3H\dot{\chi}+[m^{2}+6\xi (\dot{H}+2H^{2})]\chi-\frac{1}{a^{2}}\mathcal{D}_{i}\mathcal{D}_{i}\chi=0,
\end{equation}
where we took into account that the curvature scalar equals $R=-6(\dot{H}+2H^{2})$ for the FLRW metric and $H=\dot{a}/a$ is the Hubble parameter. As usual, an overdot denotes a derivative with respect to the cosmic 
time $t$ and we used the Coulomb gauge for the electromagnetic field, i.e., $A_{\mu}=(0,\mathbf{A})$ and ${\rm div\,}\mathbf{A}=0$. We assume that only a spatially uniform electric field is present and the magnetic field is absent, so that $\mathbf{A}=\mathbf{A}(t)$. This approximation is valid, e.g., in the kinetic coupling model with a monotonically decreasing in time coupling function
\cite{Kanno:2009,Sobol:2018}.

As usual, the charged scalar field operator can be expanded over the set of creation and annihilation operators
\begin{equation}
\label{decomposition}
\chi(t,\mathbf{x})=\int \frac{d^{3}\mathbf{k}}{(2\pi a)^{3/2}}\left[\hat{a}_{\mathbf{k}}\chi_{\mathbf{k}}(t)e^{i\mathbf{k}\cdot\mathbf{x}}
+\hat{b}_{\mathbf{k}}^{\dagger}\chi^{*}_{-\mathbf{k}}(t)e^{-i\mathbf{k}\cdot\mathbf{x}}\right],
\end{equation}
where the creation ($\hat{a}_{\mathbf{k}}^{\dagger}$, $\hat{b}_{\mathbf{k}}^{\dagger}$) and annihilation ($\hat{a}_{\mathbf{k}}$, $\hat{b}_{\mathbf{k}}$)  operators satisfy the canonical commutation relations
\begin{equation}
[\hat{a}_{\mathbf{k}},\,\hat{a}^{\dagger}_{\mathbf{p}}]=[\hat{b}_{\mathbf{k}},\,\hat{b}^{\dagger}_{\mathbf{p}}]=\delta^{3}(\mathbf{k}-\mathbf{p}),
\end{equation}
while all other commutators vanish. Substituting decomposition (\ref{decomposition}) into Eq.~(\ref{eq-scalar-2}), we obtain the equation for the mode function
\begin{equation}
\label{eq-osc}
\ddot{\chi}_{\mathbf{k}}(t)+\Omega_{\mathbf{k}}^{2}(t)\chi_{\mathbf{k}}(t)=0.
\end{equation}
Clearly, this is the equation of motion of a harmonic oscillator with time varying frequency
\begin{equation}
\label{frequency-general}
\Omega_{\mathbf{k}}^{2}(t)=m^{2}+\frac{1}{a^{2}}(\mathbf{k}-e\mathbf{A})^{2}+\Big(12\xi-\frac{9}{4}\Big)H^{2}+\Big(6\xi-\frac{3}{2}\Big)\dot{H}.
\end{equation}
In what follows we will consider only modes for which the square of frequency is positive. More precisely, we are interested in modes which are inside the Hubble horizon with physical momentum
$p_{\rm ph}\equiv |\mathbf{k}-e\mathbf{A}|/a \gtrsim H$. If $5/48<\xi<1/4$, expression (\ref{frequency-general}) is positively defined for all subhorizon modes and for any expansion rate of the universe. (Note, that $\dot{H}<0$ for any matter content of the universe satisfying the null-energy condition $\rho+P\geq 0$, where $\rho$ is the energy density and $P$ is pressure.) To avoid unnecessary complications, we 
assume that $\xi$ lies in the specified range.

As usual, we assume that the electric field was absent before the initial moment of time $t_0$ and the scalar field was in the Bunch-Davies vacuum state \cite{Bunch:1978}
\begin{equation}
\label{Bunch-Davies}
\chi_{\mathbf{k}}(t)=\frac{1}{\sqrt{2k/a}}e^{-ik\eta(t)},\qquad |k\eta(t)|\gg 1,
\end{equation}
where $\eta$ is the conformal time, $d\eta=dt/a(t)$. In terms of the latter, Eq.~(\ref{eq-osc}) takes the form
\begin{equation}
\label{eq-mode-conformal}
\left(\frac{\chi_{\mathbf{k}}}{\sqrt{a}}\right)^{\!\!\prime\prime}+\tilde{\Omega}^{2}(\eta)\frac{\chi_{\mathbf{k}}}{\sqrt{a}}=0, \qquad \tilde{\Omega}^{2}(\eta)=(\mathbf{k}-e\mathbf{A})^{2}+a^{2}m^{2}+(6\xi-1)a''/a,
\end{equation}
where prime denotes a derivative with respect to the conformal time. For $\mathbf{A}=0$, a sufficiently light field $m\lesssim H$, and $|k\eta|\approx k/(aH)\gg 1$, Eq.~(\ref{Bunch-Davies}) is a 
solution to Eq.~(\ref{eq-mode-conformal}).

In a nonzero electric field, one cannot separate the positive- and negative-frequency solutions of Eq.~(\ref{eq-osc}), i.e., the mode function contains, in general, both components and their 
relative contributions can be characterized by the Bogolyubov coefficients $\alpha_{\mathbf{k}}$ and $\beta_{\mathbf{k}}$ as follows:
\begin{equation}
\label{Bogolyubov-decomposition}
\chi_{\mathbf{k}}(t)=\frac{1}{\sqrt{2\Omega_{\mathbf{k}}(t)}}\left[\alpha_{\mathbf{k}}(t)e^{-i\Theta_{\mathbf{k}}(t)}+\beta_{\mathbf{k}}(t)e^{i\Theta_{\mathbf{k}}(t)}\right],
\end{equation}
where $\Theta_{\mathbf{k}}(t)=\int_{t_0}^{t}\Omega_{\mathbf{k}}(t')dt'$ and the coefficients satisfy the relation
\begin{equation}
\label{normalization}
|\alpha_{\mathbf{k}}(t)|^{2}-|\beta_{\mathbf{k}}(t)|^{2}=1.
\end{equation}
Initially, $\alpha_{\mathbf{k}}(t_{0})=1$, $\beta_{\mathbf{k}}(t_{0})=0$.

It is straightforward to check that Eq.~(\ref{eq-osc}) is identically satisfied if the Bogolyubov coefficients evolve according to the following system of equations:
\begin{equation}
\label{sys-alpha-beta}
\left\{
\begin{array}{l}
\dot{\alpha}_{\mathbf{k}}=\frac{\dot{\Omega}_{\mathbf{k}}}{2\Omega_{\mathbf{k}}}e^{2i\Theta_{\mathbf{k}}}\beta_{\mathbf{k}},\\
\dot{\beta}_{\mathbf{k}}=\frac{\dot{\Omega}_{\mathbf{k}}}{2\Omega_{\mathbf{k}}}e^{-2i\Theta_{\mathbf{k}}}\alpha_{\mathbf{k}}.
\end{array}
\right.
\end{equation}
Moreover, these equations ensure that the normalization condition (\ref{normalization}) is satisfied at any time as soon as it is true at the initial moment.

In view of Eq.~(\ref{normalization}), there are only three real degrees of freedom which are conveniently parametrized as follows \cite{Schmidt:1998}:
\begin{eqnarray}
\mathcal{F}_{\mathbf{k}}(t)&=&|\beta_{\mathbf{k}}|^{2},\label{func-F}\\
\mathcal{G}_{\mathbf{k}}(t)&=&\Re e\left(\alpha_{\mathbf{k}}\beta^{*}_{\mathbf{k}} e^{-2i\Theta_{\mathbf{k}}}\right),\label{func-G}\\
\mathcal{H}_{\mathbf{k}}(t)&=&\Im m\left(\alpha_{\mathbf{k}}\beta^{*}_{\mathbf{k}} e^{-2i\Theta_{\mathbf{k}}}\right).\label{func-H}
\end{eqnarray}
Equations (\ref{sys-alpha-beta}) imply the following equations of motion:
\begin{eqnarray}
\dot{\mathcal{F}}_{\mathbf{k}}&=&\frac{\dot{\Omega}_{\mathbf{k}}}{\Omega_{\mathbf{k}}}\mathcal{G}_{\mathbf{k}},\label{eq-F-1}\\
\dot{\mathcal{G}}_{\mathbf{k}}&=&\frac{\dot{\Omega}_{\mathbf{k}}}{2\Omega_{\mathbf{k}}}(1+2\mathcal{F}_{\mathbf{k}})+2\Omega_{\mathbf{k}}\mathcal{H}_{\mathbf{k}},\label{eq-G-1}\\
\dot{\mathcal{H}}_{\mathbf{k}}&=&-2\Omega_{\mathbf{k}}\mathcal{G}_{\mathbf{k}}.\label{eq-H-1}
\end{eqnarray}
It is convenient to change the variable from the canonical momentum $\mathbf{k}$ to the physical one $\mathbf{p}=(\mathbf{k}-e\mathbf{A})/a(t)$. Then
\begin{equation}
\Omega_{\mathbf{k}}\ \ \Rightarrow\ \  \omega(t,\mathbf{p})=\sqrt{m^{2}+\mathbf{p}^{2}+\Big(12\xi-\frac{9}{4}\Big)H^{2}+\Big(6\xi-\frac{3}{2}\Big)\dot{H}},
\end{equation}
\begin{equation}
\frac{\dot{\Omega}_{\mathbf{k}}}{\Omega_{\mathbf{k}}}\ \ \Rightarrow\ \ Q(t,\mathbf{p})=\frac{e\mathbf{E}\cdot \mathbf{p}-H\mathbf{p}^{2}+\Big(12\xi-\frac{9}{4}\Big)H\dot{H}
+\Big(3\xi-\frac{3}{4}\Big)\ddot{H}}{\omega^{2}(t,\mathbf{p})},
\end{equation}
where the electric field measured by a comoving observer is defined as $\mathbf{E}=-\dot{\mathbf{A}}/a$, $E^{i}=aF^{0i}$.

Finally, we rewrite Eqs.~(\ref{eq-F-1})--(\ref{eq-H-1}) in terms of the new variable. Introducing 
$\tilde{\mathcal{F}}(t,\mathbf{p})=\tilde{\mathcal{F}}(t,(\mathbf{k}-e\mathbf{A})/a)\equiv \mathcal{F}_{\mathbf{k}}(t)$
(and proceeding similarly for $\mathcal{G}$ and $\mathcal{H}$), we obtain for the time derivative
\begin{equation}
\dot{\mathcal{F}}_\mathbf{k}=\frac{\partial \tilde{\mathcal{F}}(t,\mathbf{p})}{\partial t} + \frac{\partial \tilde{\mathcal{F}}(t,\mathbf{p})}{\partial \mathbf{p}} \frac{\partial}{\partial t}\left(\frac{\mathbf{k}-e\mathbf{A}}{a}\right)=\left[\frac{\partial}{\partial t}+(e\mathbf{E}-H\mathbf{p})\frac{\partial}{\partial \mathbf{p}}\right]\tilde{\mathcal{F}}(t,\mathbf{p}).
\end{equation}
In what follows, we omit the tilde and, for the sake of brevity, introduce the quantity $\boldsymbol{\mathcal{E}}=e\mathbf{E}$.

Finally, the system of equations takes the form
\begin{eqnarray}
\left[\frac{\partial}{\partial t}+(\boldsymbol{\mathcal{E}}-H\mathbf{p})\frac{\partial}{\partial \mathbf{p}}\right]\mathcal{F}(t,\mathbf{p})&=&Q(t,\mathbf{p})\mathcal{G}(t,\mathbf{p}),\label{eq-F-2}\\
\left[\frac{\partial}{\partial t}+(\boldsymbol{\mathcal{E}}-H\mathbf{p})\frac{\partial}{\partial \mathbf{p}}\right]\mathcal{G}(t,\mathbf{p})&=&\frac{1}{2}Q(t,\mathbf{p})[1+2\mathcal{F}(t,\mathbf{p})]+2\omega(t,\mathbf{p})\mathcal{H}(t,\mathbf{p}),\label{eq-G-2}\\
\left[\frac{\partial}{\partial t}+(\boldsymbol{\mathcal{E}}-H\mathbf{p})\frac{\partial}{\partial \mathbf{p}}\right]\mathcal{H}(t,\mathbf{p})&=&-2\omega(t,\mathbf{p})\mathcal{G}(t,\mathbf{p}).\label{eq-H-2}
\end{eqnarray}
These quantum Vlasov equations describe the creation of charged scalar particles by time-dependent electric and gravitational fields. The electric current, energy density, and pressure of produced particles can 
be expressed in terms of the kinetic functions $\mathcal{F}$, $\mathcal{G}$, and $\mathcal{H}$. However, as usual in relativistic quantum field systems, it turns out that there are UV divergences. In the next section, we study the UV asymptotics of the kinetic functions and apply the renormalization procedure in order to render physical observables finite.

\section{Renormalization}
\label{sec-UV}

Since UV divergences of the electric current $j^{\mu}$ and energy-momentum tensor $T^{\mu\nu}$ are connected with the behavior of the corresponding integrands at large momentum, our approach to divergences and
the subsequent renormalization procedure is standard. First we determine the UV behavior of the kinetic functions $\mathcal{F}(t,\mathbf{p})$, $\mathcal{G}(t,\mathbf{p})$, and $\mathcal{H}(t,\mathbf{p})$ at
$|\mathbf{p}| \to \infty$ as a series in 
inverse powers of $\epsilon_p=\sqrt{\mathbf{p}^2+m^2}$ (an expansion in inverse powers of $|\mathbf{p}|$ is inconvenient because the integration over momentum will lead to infrared divergences). Since $j^{\mu}$ and
$T^{\mu\nu}$ are expressed via integrals of the kinetic functions, the found expansions allow us to find the divergent terms of the electric current and energy-momentum tensor.
Finally, we determine counterterms which cancel the divergences.

\subsection{Asymptotic behavior of kinetic functions}
\label{subsec-UV}

In order to find the asymptotic UV behavior of the kinetic functions which satisfy the system of equations (\ref{eq-F-2})--(\ref{eq-H-2}), we represent them as a series in inverse powers of $\epsilon_p$ at
large momentum. We assume that each of these series starts with a certain negative power, i.e., $\mathcal{F}\sim \epsilon_p^{n_{1}}$, $\mathcal{G}\sim \epsilon_p^{n_{2}}$, and $\mathcal{H}\sim \epsilon_p^{n_{3}}$ with 
negative integers $n_{1}$, $n_{2}$, and $n_{3}$. Then Eq.~(\ref{eq-F-2}) implies $n_{1}=n_{2}$, while Eq.~(\ref{eq-H-2}) gives $n_{3}=n_{2}+1$. Finally, Eq.~(\ref{eq-G-2}) produces $n_{3}=-1$. Substituting expansions 
(\ref{dec-f})--(\ref{dec-h}) in Eqs.~(\ref{eq-F-2})--(\ref{eq-H-2}) and collecting terms with the same power of $|\mathbf{p}|$, we obtain the system of Eqs.~(\ref{eq-h-1})--(\ref{eq-f-4}) which determines the 
leading terms in the UV power expansions of $\mathcal{F}$, $\mathcal{G}$, and $\mathcal{H}$. 

Using the results obtained in Appendix~\ref{app-UV}, we find the following asymptotic expansions for the kinetic functions:
\begin{multline}
\label{f-decomp-eps}
\mathcal{F}(t,\mathbf{p})=\frac{H^{2}}{16\epsilon_{p}^{2}}-\frac{H\mathbf{v}\cdot\boldsymbol{\mathcal{E}}}{8\epsilon_{p}^{3}}
+\frac{(\mathbf{v}\cdot\boldsymbol{\mathcal{E}})^{2}}{16\epsilon_{p}^{4}}+\\
+(1-6\xi)\frac{H\ddot{H}+7H^{2}\dot{H}+6H^{4}}{16\epsilon_{p}^{4}}+\frac{\dot{H}^{2}+2H^{2}\dot{H}+\frac{3}{4}H^{4}-8 m^{2}H^{2}}{64\epsilon_{p}^{4}}+\mathcal{O}(\epsilon_{p}^{-5}),
\end{multline}
\begin{multline}
\label{g-decomp-eps}
\mathcal{G}(t,\mathbf{p})=-\frac{\dot{H}+H^{2}}{8\epsilon_{p}^{2}}+\frac{\mathbf{v}\cdot\big(\dot{\boldsymbol{\mathcal{E}}}+3H\boldsymbol{\mathcal{E}}\big)}{8\epsilon_{p}^{3}}+\frac{\boldsymbol{\mathcal{E}}^{2}-3(\mathbf{v}\cdot\boldsymbol{\mathcal{E}})^{2}}{8\epsilon_{p}^{4}}-\\
-(1-6\xi)\frac{\dddot{H}+8\dot{H}^{2}+10H\ddot{H}+42H^{2}\dot{H}+20H^{4}}{16\epsilon_{p}^{4}}-\\
-\frac{2\dot{H}^{2}+3H^{2}\dot{H}+H^{4}-32 m^{2}H^{2}-8m^{2}\dot{H}}{64\epsilon_{p}^{4}}+\mathcal{O}(\epsilon_{p}^{-5}),
\end{multline}
\begin{equation}
\label{h-decomp-eps}
\mathcal{H}(t,\mathbf{p})=\frac{H}{4\epsilon_{p}}-\frac{\mathbf{v}\cdot\boldsymbol{\mathcal{E}}}{4\epsilon_{p}^{2}}
+\frac{(1-6\xi)\big(\ddot{H}+7H\dot{H}+6H^{3}\big)-2m^{2}H}{8\epsilon_{p}^{3}}+\mathcal{O}(\epsilon_{p}^{-4}),
\end{equation}
where $\mathbf{v}=\mathbf{p}/\epsilon_{p}$ is the particle's velocity.

\subsection{Electric current}
\label{subsec-current}

The electric current is defined as the vacuum expectation value of the corresponding electric current operator
\begin{equation}
j^{\nu}\equiv \left<\frac{\partial\mathcal{L}_{\rm ch}}{\partial A_{\nu}}\right>=ie\langle\chi^{\dagger}\overleftrightarrow{\mathcal{D}}^{\nu}\chi\rangle,
\end{equation}
where $\overleftrightarrow{\mathcal{D}}_{\mu}=\overrightarrow{\mathcal{D}}_{\mu}-\overleftarrow{\mathcal{D}}^{*}_{\mu}$ and arrows indicate the position where (to the left 
or to the right) the covariant derivative acts. 
Then it is straightforward to derive from Eqs.~(\ref{lagr-EM-gen}) and (\ref{lagr-chi}) the equation describing the evolution of electric field \cite{Gorbar:2019}
\begin{equation}
\label{Maxwell-eq}
\dot{\boldsymbol{\mathcal{E}}}+2H\boldsymbol{\mathcal{E}}=\frac{e}{a}\mathbf{j} + \text{(source)},
\end{equation}
where the last term corresponds to a source of electric field which is due to the term $\mathcal{L}_{\rm int}$ in Eq.~(\ref{lagr-EM-gen}). Because of the space homogeneity, the charge density vanishes (scalar particles are produced in pairs together with their antiparticles everywhere in space). Using the definition of electric current
\begin{equation}
\mathbf{j}=ie\left<\chi^{\dagger}\boldsymbol{\nabla}\chi-\chi\boldsymbol{\nabla}\chi^{\dagger}-2ie \mathbf{A} \chi^{\dagger}\chi\right>,
\end{equation}
decomposition (\ref{decomposition}), and the properties of the creation and annihilation operators, it is straightforward to derive
\begin{equation}
\mathbf{j}=-\frac{2e}{a^{3}}\int\frac{d^{3}\mathbf{k}}{(2\pi)^{3}}(\mathbf{k}-e\mathbf{A})|\chi_{\mathbf{k}}|^{2}.
\end{equation}
Using Eqs.(\ref{Bogolyubov-decomposition}) and (\ref{func-F})--(\ref{func-G}), we obtain
\begin{equation}
\label{current-non-renormalized}
\mathbf{j}=-ae\int\frac{d^{3}\mathbf{k}}{(2\pi a)^{3}}\frac{(\mathbf{k}-e\mathbf{A})}{a \Omega_{\mathbf{k}}}\left[1+2\mathcal{F}_{\mathbf{k}}+2\mathcal{G}_{\mathbf{k}}\right]=-2ae \int\frac{d^{3}\mathbf{p}}{(2\pi)^{3}}\mathbf{p}\frac{\mathcal{F}(t,\mathbf{p})+\mathcal{G}(t,\mathbf{p})}{\omega(t,\mathbf{p})}.
\end{equation}

Since the kinetic functions decrease only as $\sim p^{-2}$ for $p\to\infty$, current (\ref{current-non-renormalized}) is divergent in the UV region and has to be renormalized. Terms of 
order $p^{-3}$ also lead to divergences. Collecting the appropriate terms in Eqs.~(\ref{f-decomp-eps}) and (\ref{g-decomp-eps}), we derive
\begin{equation}
\frac{\mathcal{F}(t,\mathbf{p})+\mathcal{G}(t,\mathbf{p})}{\omega(t,\mathbf{p})}=-\frac{2\dot{H}+H^{2}}{16\epsilon_{p}^{3}}+\frac{\mathbf{v}\cdot\big(\dot{\boldsymbol{\mathcal{E}}}+2H\boldsymbol{\mathcal{E}}\big)}{8\epsilon_{p}^{4}}+\mathcal{O}(\epsilon_{p}^{-5}).
\end{equation}
The first term corresponds to the particle production due to the universe expansion and could lead to a severe (linear) UV divergence of the current. However, its contribution vanishes because it is an even 
function of momentum and its integral with $\mathbf{p}$ over momentum vanishes. This result is a consequence of the fact that the gravitational particle 
production is not sensitive to their charge. In contrast 
to that, the second term is an odd function of momentum. It behaves as $\sim p^{-4}$ at infinity and leads to a logarithmic UV divergence of the electric current. All other terms are at least $\sim p^{-5}$ and, therefore,
produce a regular contribution $\mathbf{j}_{\rm reg}$. Since the dimensional regularization preserves the gauge invariance and the covariance with respect to differentiable coordinate transformations, we find it
very convenient. Using Eq.~(\ref{int-vi-vj}), we obtain the divergent part of the current
\begin{equation}
\label{j-div}
e\mathbf{j}_{\rm div}=-\frac{ae^{2}}{48\pi^{2}} (\dot{\boldsymbol{\mathcal{E}}}+2H\boldsymbol{\mathcal{E}}) \left(\Delta_{\varepsilon}+\ln\frac{\mu_{r}^{2}}{m^{2}}\right),
\end{equation}
where $\Delta_{\varepsilon}$ is the divergent contribution extracted in the $\overline{\rm MS}$ scheme, see Eq.~(\ref{Upsilon}), and $\mu_{r}$ is an arbitrary energy scale arising in the course of regularization, 
see Appendix~\ref{app-dimensional}. The regular part of the electric current reads
\begin{equation}
\label{j-reg}
e\mathbf{j}_{\rm reg}=-2ae^{2} \int\frac{d^{3}\mathbf{p}}{(2\pi)^{3}}\mathbf{p}\left[\frac{\mathcal{F}(t,\mathbf{p})+\mathcal{G}(t,\mathbf{p})}{\omega(t,\mathbf{p})}-\frac{\mathbf{p}\cdot(\dot{\boldsymbol{\mathcal{E}}}+2H \boldsymbol{\mathcal{E}})}{8(p^{2}+m^{2})^{5/2}}\right],
\end{equation}
where we subtracted from the integrand the terms which lead to UV divergences.

\subsection{Energy-momentum tensor of produced particles}
\label{subsec-EMT}

The expansion of the FLRW universe is described by the Friedmann equations
\begin{equation}
\label{eq-Friedmann}
H^{2}=\frac{1}{3M_{p}^{2}}\rho^{({\rm tot})},\qquad \dot{H}=-\frac{1}{2M_{p}^{2}}(\rho^{({\rm tot})}+P^{({\rm tot})}),
\end{equation}
sourced by the total energy density $\rho^{({\rm tot})}$ and pressure $P^{({\rm tot})}$ of the matter fields filling the universe. They can be determined from the corresponding energy-momentum tensor.
By definition,
\begin{equation}
T^{\mu\nu}_{\rm mat}=-\frac{2}{\sqrt{-g}}\left<\frac{\delta S_{\rm mat}}{\delta g_{\mu\nu}}\right>=T^{\mu\nu}_{\rm inf}+T^{\mu\nu}_{\rm EM}+T^{\mu\nu}_{\chi},
\end{equation}
where the first term describes the contribution of the inflaton field driving inflation, the second corresponds to the contribution of the electromagnetic field, and the last 
term quantifies the contribution of the charged scalar field whose explicit expression is given by Eq.~(\ref{T-mu-nu-chi}).

The 00 component of the energy-momentum tensor determines the energy density of the matter fields while pressure can be expressed through trace $T=T^{\mu}_{\mu}$ as $P=(\rho-T)/3$. Then the contributions of the charged 
scalar field to $\rho$ and $T$ are
\begin{equation}
\rho_{\chi}=\langle|\mathcal{D}_{0}\chi|^{2}+\frac{1}{a^{2}}|\mathcal{D}_{i}\chi|^{2}+(m^{2}+6H^{2}\xi)|\chi|^{2}+6H\xi\partial_{0}|\chi|^{2}-\frac{2\xi}{a^{2}}\partial_{i}^{2}|\chi|^{2}\rangle.
\label{rho-chi-1}
\end{equation}
\begin{equation}
\label{trace-T-chi}
T_{\chi}=\left<2(6\xi-1)\left[(\mathcal{D}_{\lambda}\chi)^{\dagger}(\mathcal{D}^{\lambda}\chi)+\xi R|\chi|^{2}\right]+4(1-3\xi)m^{2}|\chi|^{2}\right>.
\end{equation}
In the derivation of $T_{\chi}$ the Klein-Gordon-Fock equation (\ref{eq-scalar-2}) was used.
It is remarkable that the trace $T_{\chi}$ vanishes in the massless case $m=0$ and for $\xi=1/6$. This result is due to the conformal symmetry of the classical action. Therefore, $\xi=1/6$ 
is known as the conformal coupling.

Using Eq.~(\ref{decomposition}), the vacuum expectation values (\ref{rho-chi-1}) and (\ref{trace-T-chi}) are calculated in Appendix~\ref{app-EMT}. Expressing them in terms of the kinetic functions, we obtain
\begin{equation}
\label{rho-chi}
\rho_{\chi}=\int\frac{d^{3}\mathbf{p}}{(2\pi)^{3}}\Big\{\omega(t,\mathbf{p})\big[2\mathcal{F}(t,\mathbf{p})+1\big]-3H(1-4\xi)\mathcal{H}(t,\mathbf{p})
+\frac{(9-48\xi)H^{2}+(3-12\xi)\dot{H}}{2\omega(t,\mathbf{p})}\Big[\mathcal{F}(t,\mathbf{p})+\mathcal{G}(t,\mathbf{p})+\frac{1}{2}\Big] \Big\}.
\end{equation}
\begin{equation}
\label{T-chi}
T_{\chi}=\int\frac{d^{3}\mathbf{p}}{(2\pi)^{3}}\bigg\{-2(6\xi-1)\big[2\omega(t,\mathbf{p})\mathcal{G}(t,\mathbf{p})+3H\mathcal{H}(t,\mathbf{p})\big]
+\frac{2m^{2}-3(6\xi-1)\dot{H}}{\omega(t,\mathbf{p})}\Big[\mathcal{F}(t,\mathbf{p})+\mathcal{G}(t,\mathbf{p})+\frac{1}{2}\Big]\bigg\}.
\end{equation}
These expressions contain divergent contributions which are defined by Eqs.~(\ref{energy-density-divergence}) and (\ref{trace-divergence}).
Using the dimensional regularization in the $\overline{\rm MS}$ scheme, we obtain
\begin{equation}
\rho_{\chi}=\rho_{\chi}^{\rm div}+\rho_{\chi}^{\rm reg}, \qquad T_{\chi}=T_{\chi}^{\rm div}+T_{\chi}^{\rm reg},
\end{equation}
where the divergent parts read as
\begin{equation}
\label{rho-div}
\rho_{\chi}^{\rm div}=\bigg\{-\frac{m^{4}}{32\pi^{2}}+\frac{m^{2}H^{2}}{16\pi^{2}}(1-6\xi)-\frac{2H\ddot{H}-\dot{H}^{2}+6H^{2}\dot{H}}{32\pi^{2}}(1-6\xi)^{2}+\frac{\boldsymbol{\mathcal{E}}^{2}}{96\pi^{2}}\bigg\}\left(\Delta_{\varepsilon}+\ln\frac{\mu_{r}^{2}}{m^{2}}\right),
\end{equation}
\begin{equation}
\label{T-div}
T_{\chi}^{\rm div}=\bigg\{
-\frac{m^{4}}{8\pi^{2}}+\frac{m^{2}(\dot{H}+2H^{2})}{8\pi^{2}}(1-6\xi)-\frac{\dddot{H}+4\dot{H}^{2}+7H\ddot{H}+12H^{2}\dot{H}}{16\pi^{2}}(1-6\xi)^{2}
\bigg\}\left(\Delta_{\varepsilon}+\ln\frac{\mu_{r}^{2}}{m^{2}}\right),
\end{equation}
while the regular parts are given by
\begin{eqnarray}
\label{rho-chi-reg-1}
\rho_{\chi}^{\rm reg}&=&-\frac{3m^{4}}{64\pi^{2}}-\frac{m^{2}H^{2}}{16\pi^{2}}(1-6\xi)+\int\frac{d^{3}\mathbf{p}}{(2\pi)^{3}}\Big\{2\omega(t,\mathbf{p})\big[\mathcal{F}(t,\mathbf{p})+1/2\big]-3H(1-4\xi)\mathcal{H}(t,\mathbf{p})+\nonumber\\
&+&\frac{(9-48\xi)H^{2}+(3-12\xi)\dot{H}}{2\omega(t,\mathbf{p})}\big[\mathcal{F}(t,\mathbf{p})+\mathcal{G}(t,\mathbf{p})+1/2\big]-\epsilon_{p}-\frac{H^{2}}{2\epsilon_{p}}(1-6\xi)-\nonumber\\
&-&\frac{(\mathbf{v}\cdot\boldsymbol{\mathcal{E}})^{2}}{8\epsilon_{p}^{3}}-\frac{H^{2}m^{2}}{2\epsilon_{p}^{3}}(1-6\xi)+\frac{2H\ddot{H}-\dot{H}^{2}+6H^{2}\dot{H}}{8\epsilon_{p}^{3}}(1-6\xi)^{2} \Big\},
\end{eqnarray}
\begin{eqnarray}
\label{T-chi-reg-1}
T_{\chi}^{\rm reg}&=&-\frac{m^{4}}{8\pi^{2}}-\frac{m^{2}(\dot{H}+H^{2})}{8\pi^{2}}(1-6\xi)+\int\frac{d^{3}\mathbf{p}}{(2\pi)^{3}}\bigg\{2(1-6\xi)\big[2\omega(t,\mathbf{p})\mathcal{G}(t,\mathbf{p})+3H\mathcal{H}(t,\mathbf{p})\big]+\nonumber\\
&+&\big[2m^{2}+3(1-6\xi)\dot{H}\big]\frac{1}{\omega(t,\mathbf{p})}\big[\mathcal{F}(t,\mathbf{p})+\mathcal{G}(t,\mathbf{p})+1/2\big]-\frac{m^{2}}{\epsilon_{p}}-\frac{\dot{H}+H^{2}}{\epsilon_{p}}(1-6\xi)-\nonumber\\
&-&\frac{m^{2}(2\dot{H}+3H^{2})}{2\epsilon_{p}^{3}}(1-6\xi)+\frac{\dddot{H}+4\dot{H}^{2}+7H\ddot{H}+12H^{2}\dot{H}}{4\epsilon_{p}^{3}}(1-6\xi)^{2}
\bigg\}.
\end{eqnarray}

We would like to emphasize once again that the use of the dimensional regularization allowed us to extract the divergent contributions preserving the gauge invariance as well as the general covariance. Therefore, these 
contributions can be canceled by covariant counterterms 
in the action. These counterterms will be determined in the next subsection. It is important to note that not every regularization scheme leads to a covariant form of the counterterms (e.g., a simple momentum cutoff does not).
A relevant instructive example in the Minkowski spacetime was considered in Ref.~\cite{Akhmedov:2002}.

\subsection{Counterterms}
\label{subsec-counterterms}

In the previous subsections we extracted the UV divergent parts of such physical observables as the electric current, energy density, and pressure. Here we look for the appropriate counterterms to cancel these 
divergences and render the corresponding physical quantities finite.

The divergent parts (\ref{j-div}), (\ref{rho-div}), and (\ref{T-div}) depend on the renormalization scale $\mu_{r}$. In an inflating universe, a natural choice would be $\mu_{r}=H_{e}$, where $H_{e}$ is the 
Hubble parameter at the end of inflation when the Schwinger pair production takes place. However, the choice $\mu_{r}=m$ is more convenient because it allows us to use the values of the electric charge $e$ 
and other coupling constants determined experimentally. In such a case, their running to the actual energy scale during inflation is automatically encoded in the electric current and energy-momentum tensor.

Let us first deal with the divergence in the electric current (\ref{j-div}). Since its dependence on the electric field is the same as in the left-hand side of the Maxwell equation (\ref{Maxwell-eq}), 
the counterterm has the usual form
\begin{equation}
\label{counterterm-Z3}
S_{Z_{3}}=-\int d^{4}x \sqrt{-g}\,\frac{Z_{3}-1}{4}F_{\mu\nu}F^{\mu\nu}.
\end{equation}
It is easy to show that the choice
\begin{equation}
\label{Z_3}
Z_{3}=1-\frac{e^{2}}{48\pi^{2}}\Delta_{\varepsilon}
\end{equation}
cancels the divergent contribution in the Maxwell equation. Here $\Delta_{\varepsilon}$ is defined in Eq.~(\ref{Upsilon}). Indeed, instead of Eq.~(\ref{Maxwell-eq}), we obtain the following equation for the electric field:
\begin{equation}
\left[1+(Z_{3}-1)\right](\dot{\boldsymbol{\mathcal{E}}}+2H\boldsymbol{\mathcal{E}})=\frac{e}{a}(\mathbf{j}_{\rm div}+\mathbf{j}_{\rm reg})+ \text{(source)}.
\end{equation}
Since the counterterm contribution on the left-hand side is exactly the same as the divergent part of current (\ref{j-div}) on the right-hand side, the resulting equation contains only finite quantities.
The same result was obtained in Ref.~\cite{Banyeres:2018} in the case of a constant electric field in de Sitter spacetime and coincides with the standard textbook expression for the electromagnetic field renormalization 
constant in scalar electrodynamics~\cite{Srednicki-book}.

Thus, a system of equations describing the production of charged scalar particles and their backreaction on the electric field evolution consists of the three coupled kinetic
equations (\ref{eq-F-2})--(\ref{eq-H-2}) and the Maxwell equation with renormalized current
\begin{equation}
\label{Maxwell-eq-renorm}
\dot{\boldsymbol{\mathcal{E}}}+2H\boldsymbol{\mathcal{E}}=-2e^{2}\int\frac{d^{3}\mathbf{p}}{(2\pi)^{3}}\mathbf{p}\left[\frac{\mathcal{F}(t,\mathbf{p})+\mathcal{G}(t,\mathbf{p})}{\omega(t,\mathbf{p})}-\frac{\mathbf{p}\cdot(\dot{\boldsymbol{\mathcal{E}}}+2H \boldsymbol{\mathcal{E}})}{8(p^{2}+m^{2})^{5/2}}\right]+\text{(source)}.
\end{equation}

The backreaction of the electromagnetic field and charged particles on the universe expansion should be taken into account too. For this, we should renormalize the energy density and the pressure of produced 
scalar particles obtained in Sec.~\ref{subsec-EMT}. Let us start with the energy density (\ref{rho-div}). Clearly, there is only one term which depends on the electric field and which is exactly canceled by the $Z_3$ term 
coming from the electromagnetic energy density
\begin{equation}
\label{rho-Z3}
\delta\rho_{\rm EM}^{Z_{3}}=(Z_{3}-1)\frac{\mathbf{E}^{2}}{2}=-\frac{\boldsymbol{\mathcal{E}}^{2}}{96\pi^{2}}\Delta_{\varepsilon},
\end{equation}
which is the 00 component of the effective energy momentum tensor (\ref{EMT-Z3}). All remaining divergent terms can be removed by additional counterterms in the Lagrangian. Let us seek them in the following simple form:
\begin{equation}
\label{grav-ct}
S_{\rm ct}=\int d^{4}x \sqrt{-g} \left[a_{1} +a_{2}R+a_{3} R^{2}\right],
\end{equation}
where $a_1$, $a_2$, and $a_{3}$ has to be determined by requiring the cancellation of all divergent terms in the energy density of the scalar field. While the first term corresponds to a 
renormalization of the cosmological constant, the second defines a correction to the Planck mass. The third term is absent in the original gravitational action and its emergence demonstrates the nonrenormalizability of the 
Einstein gravity. Its appearance in typical matter loop radiative corrections is one of the main reasons why Starobinsky introduced his well-known $R^{2}$-term in the action
\cite{Starobinsky:1980}.

Varying action (\ref{grav-ct}) with respect to the metric, we obtain the following energy-momentum tensor:
\begin{equation}
\label{EMT-ct}
T^{\mu\nu}_{\rm ct}=-\frac{2}{\sqrt{-g}}\frac{\delta S_{\rm ct}}{\delta g_{\mu\nu}}=-a_{1}g^{\mu\nu} +a_{2}(2R^{\mu\nu}-Rg^{\mu\nu})+4a_{3}\left[RR^{\mu\nu}-\frac{1}{4}R^{2}g^{\mu\nu}-(\nabla^{\mu}\nabla^{\nu}
-g^{\mu\nu}\nabla_{\lambda}\nabla^{\lambda})R\right]
\end{equation}
whose 00 component defines the energy density. Calculating it for a spatially flat FLRW metric, we find
\begin{equation}
\rho_{\rm ct}=-a_{1}+6H^{2}a_{2}-36a_{3}(2H\ddot{H}-\dot{H}^{2}+6H^{2}\dot{H}),
\end{equation}
which has exactly the same functional dependence on $H$ and its derivatives as the divergent terms in the energy density of the scalar field (\ref{rho-div}). Then the condition
$\rho_{\chi}^{\rm div}+\rho_{\rm ct}=0$ determines the following renormalization constants:
\begin{eqnarray}
a_{1}&=&-\frac{m^{4}}{32\pi^{2}}\Delta_{\varepsilon},\label{a-1}\\
a_{2}&=&\left(\xi-\frac{1}{6}\right)\frac{m^{2}}{16\pi^{2}}\Delta_{\varepsilon},\label{a-2}\\
a_{3}&=&-\left(\xi-\frac{1}{6}\right)^{2}\frac{1}{32\pi^{2}}\Delta_{\varepsilon}.\label{a-3}
\end{eqnarray}
These expressions are the same as in the usual 1-loop renormalization in the Green's functions formalism (see, e.g., Refs.~\cite{Christensen:1976,Birrell-Davies}). In addition, there is also a counterterm proportional to
$C_{\mu\nu\alpha\beta}C^{\mu\nu\alpha\beta}$, where $C_{\mu\nu\alpha\beta}$ is the Weyl tensor. However, for the FLRW metric, which is conformally flat, this tensor 
identically vanishes. This is why such a counterterm is not needed in our case. 

Finally, we check that counterterms (\ref{grav-ct}) cancel also the divergence in the trace of the energy momentum tensor (\ref{T-div}). The trace of the effective energy-momentum tensor of 
counterterms (\ref{EMT-ct}) equals
\begin{equation}
T_{\rm ct}=g_{\mu\nu}T^{\mu\nu}_{\rm ct}=-4a_{1}-2R a_{2}+12 a_{3}\nabla_{\lambda}\nabla^{\lambda}R.
\end{equation}
Taking into account that $R=-6(\dot{H}+2H^{2})$, $\nabla_{\lambda}\nabla^{\lambda}R=\ddot{R}+3H\dot{R}=-6(\dddot{H}+4\dot{H}^{2}+7H\ddot{H}+12H^{2}\dot{H})$, and using Eqs.~(\ref{a-1})--(\ref{a-3}), we find
\begin{equation}
T_{\rm ct}=\Big\{\frac{m^{4}}{8\pi^{2}}-(1-6\xi)\frac{m^{2}(\dot{H}+2H^{2})}{8\pi^{2}}+(1-6\xi)^{2}\frac{\dddot{H}+4\dot{H}^{2}+7H\ddot{H}+12H^{2}\dot{H}}{16\pi^{2}} \Big\}\Delta_{\varepsilon}.
\end{equation}
Comparing this expression with Eq.~(\ref{T-div}), we conclude that the sum $T_{\chi}^{\rm div}+T_{\rm ct}$ identically vanishes. Thus, all divergences are canceled and no new counterterms are needed.

\section{Kinetic coupling model}
\label{sec-kinetic-coupling}

Having determined in the previous subsection the renormalized equations for the electric current and energy-momentum tensor, we could apply them to the study of inflationary magnetogenesis. In this section, we
consider a specific model which generates strong electric fields during inflation. In this model, the electromagnetic field is coupled to the inflaton field through a 
modified kinetic term
\begin{equation}
\label{lagr-kinetic-coupling}
\mathcal{L}_{\rm EM}=-\frac{1}{4}f^{2}(\phi)F_{\mu\nu}F^{\mu\nu}
\end{equation}
with coupling function $f(\phi)$ monotonously decreasing in time. It has to be always larger than unity in order to avoid the strong coupling regime \cite{Demozzi:2009} and tend to the value $f=1$ at the end of 
preheating (when the inflaton stops in the minimum of its potential). We would like to mention that the Lagrangian (\ref{lagr-kinetic-coupling}) cannot be represented in the form (\ref{lagr-EM-gen}); however, the corresponding Maxwell equation for $\boldsymbol{\mathcal{E}}$ has exactly 
form (\ref{Maxwell-eq}) with the redefinition $e\to e_{\rm eff}=e/f$ \cite{Gorbar:2019}. Here $e_{\rm eff}$ is the effective charge of scalar particles in the kinetic coupling model, the corresponding 
effective electric field is $E_{\rm eff}=fE$, while the quantity $\mathcal{E}=eE=e_{\rm eff}E_{\rm eff}$ is the same as in the absence of the kinetic coupling with the inflaton field when $f=1$.

The full action is given by Eq.~(\ref{action}) with counterterms (\ref{counterterm-Z3}) and (\ref{grav-ct}). For simplicity, we consider the conformal coupling $\xi=1/6$ in this and the next sections.
As was shown in Ref.~\cite{Sobol:2018}, the equation for the electric energy density has to be supplied with a boundary term which describes the quantum-to-classical transition of 
the electromagnetic modes crossing the Hubble horizon during inflation. 
Let us assume that in some large space region an almost uniform electric field is being generated during inflation. Projecting Eq.~(\ref{Maxwell-eq-renorm}) on the direction of this field, we finally get \cite{Gorbar:2019}
\begin{equation}
\label{eq-for-E}
\dot{\mathcal{E}}+2H\mathcal{E}+2\frac{\dot{f}}{f}\mathcal{E}=\frac{e_{\rm eff}}{a}j_{\parallel}+\frac{e^{2}_{\rm eff} H^{3}}{4\pi^{2}\mathcal{E}}\left[H^{2}+\left(\dot{f}/f\right)^{2}\right].
\end{equation}
Here, the last term on the left-hand side is the source of the electric field which was not specified in Eqs.~(\ref{Maxwell-eq}) and (\ref{Maxwell-eq-renorm}) while the last term on the right-hand side is the above-mentioned boundary term. It is worth noting that the latter term is essential for the generation of the electric field during inflation but has to be excluded from Eq.~(\ref{eq-for-E}) after the end of inflation when no new modes cross the Hubble horizon.

The longitudinal component of the electric current with subtracted UV divergence equals
\begin{equation}
j_{\parallel}=-2ae_{\rm eff}\int \frac{d^{3}\mathbf{p}}{(2\pi)^{3}} p_{\parallel} \left[\frac{\mathcal{F}(t,\mathbf{p})+\mathcal{G}(t,\mathbf{p})}{\omega(t,\mathbf{p})}-\frac{p_{\parallel}(\dot{\mathcal{E}}+2H \mathcal{E})}{8(p^{2}+m^{2})^{5/2}}\right].
\end{equation}
The transverse component of the electric current identically vanishes because the kinetic functions $\mathcal{F}$ and $\mathcal{G}$ as well as the counterterm are even functions of transverse momentum and their integrals with factor $\mathbf{p}_{\perp}$ vanish.

The evolution of the inflaton field is defined by the equation
\begin{equation}
\label{eq-for-inflaton}
\ddot{\phi}+3H\dot{\phi}+\frac{dV}{d\phi}=\frac{f(\phi)f'(\phi)}{e^{2}}\mathcal{E}^{2}.
\end{equation}
The universe expansion is governed by the Friedmann equations
\begin{equation}
\label{eq-Friedmann-1}
H^{2}=\frac{1}{3M_{p}^{2}}\rho^{({\rm tot})}=\frac{1}{3M_{p}^{2}}\left(\frac{1}{2}\dot{\phi}^{2}+V(\phi)+\frac{f^{2}}{2e^{2}}\mathcal{E}^{2}+\rho_{\chi}^{\rm reg}\right),
\end{equation}
\begin{equation}
\label{eq-Friedmann-2}
\dot{H}=-\frac{1}{2M_{p}^{2}}(\rho^{({\rm tot})}+P^{({\rm tot})})=-\frac{1}{2M_{p}^{2}}\left(\dot{\phi}^{2}+\frac{2f^{2}}{3e^{2}}\mathcal{E}^{2}+\frac{4\rho_{\chi}^{\rm reg}-T_{\chi}^{\rm reg}}{3}\right),
\end{equation}
where $\rho_{\chi}^{\rm reg}$ and $T_{\chi}^{\rm reg}$ are the regular parts of the energy density and the trace of the energy-momentum tensor of charged scalar particles. In the case of conformal coupling $\xi=1/6$, expressions (\ref{rho-chi-reg-1})--(\ref{T-chi-reg-1}) are significantly simplified and take the form
\begin{multline}
\rho_{\chi}^{{\rm reg},\xi=1/6}=-\frac{3m^{4}}{64\pi^{2}}+\int\frac{d^{3}\mathbf{p}}{(2\pi)^{3}}\Big\{2\omega(t,\mathbf{p})\big[\mathcal{F}(t,\mathbf{p})+1/2\big]-H\mathcal{H}(t,\mathbf{p})+\\
+\frac{H^{2}+\dot{H}}{2\omega(t,\mathbf{p})}\big[\mathcal{F}(t,\mathbf{p})+\mathcal{G}(t,\mathbf{p})+1/2\big]-\epsilon_{p}-\frac{(\mathbf{p}\cdot\boldsymbol{\mathcal{E}})^{2}}{8\epsilon_{p}^{5}}\Big\},\label{rho-chi-reg}
\end{multline}
\begin{equation}
\label{T-chi-reg}
T_{\chi}^{{\rm reg},\xi=1/6}=-\frac{m^{4}}{8\pi^{2}}+m^{2}\int\frac{d^{3}\mathbf{p}}{(2\pi)^{3}}\Big\{\frac{2}{\omega(t,\mathbf{p})}\big[\mathcal{F}(t,\mathbf{p})+\mathcal{G}(t,\mathbf{p})+1/2\big]-\frac{1}{\epsilon_{p}}\Big\}.
\end{equation}

In order to specify the inflaton dynamics, we consider the Starobinsky model of inflation \cite{Starobinsky:1980} with potential
\begin{equation}
\label{Starobinsky-pot}
V(\phi)=\frac{3\mu^{2}M_{p}^{2}}{4}\left[1-\exp\left(-\sqrt{\frac{2}{3}}\frac{\phi}{M_{p}}\right)\right]^{2},
\end{equation}
where $\mu\approx 1.3\times 10^{-5}\,M_{p}$. It belongs to the class of plateau models favored by the Planck Collaboration observations \cite{Planck:2018inf}.
It is worth noting that we do not consider the emergence of potential (\ref{Starobinsky-pot}) from the $R^{2}$ model and use it solely for illustrative purposes. Any other flat potential would apply.

The simplest choice for the coupling function $f(\phi)$ is that of the Ratra model \cite{Ratra:1992}
\begin{equation}
\label{ratra-function}
f(\phi)=\exp\left(\beta \frac{\phi}{M_{p}}\right),
\end{equation}
where $\beta$ is a free dimensionless coupling parameter which takes values in the range $5 \lesssim \beta\lesssim 15$.

The initial value of the inflaton field has to be chosen so that to provide at least 50--60 e-foldings of inflation. For the Starobinsky model, the corresponding value is
\begin{equation}
\label{init-inflaton}
\phi_{0}\approx\sqrt{\frac{3}{2}} M_{p}\ln \frac{4 N}{3}\simeq (\text{5.1--5.4})\,M_{p}, \quad\text{for} \ \ N=\text{50--60}.
\end{equation}
The initial value of the inflaton time derivative can be found from the Friedmann equation (\ref{eq-Friedmann-1}) and the scalar field equation (\ref{eq-for-inflaton}) in the slow-roll approximation
\begin{equation}
\label{initial-derivative}
\dot{\phi}_{0}\approx -\frac{V'(\phi_{0})M_{p}}{\sqrt{3V(\phi_{0})}}=-\sqrt{\frac{2}{3}}\mu M_{p} \exp\left(-\sqrt{\frac{2}{3}}\frac{\phi_{0}}{M_{p}}\right)\approx -\frac{\mu M_{p}}{2N}\sqrt{\frac{3}{2}}.
\end{equation}
Here we used the fact that the initial value of the electric field equals zero. The kinetic functions satisfy Eqs.~(\ref{eq-F-2})--(\ref{eq-H-2}) where 
\begin{equation}
\label{Q-1/6}
Q(t,\mathbf{p})=Q^{\xi=1/6}(t,\mathbf{p})=\frac{\mathcal{E} p_{\parallel}-H\mathbf{p}^{2}-\frac{1}{4}H\dot{H}-\frac{1}{4}\ddot{H}}{m^{2}+\mathbf{p}^{2}-\frac{1}{4}H^{2}-\frac{1}{2}\dot{H}}.
\end{equation}
The initial values of all kinetic functions are equal to zero. In the next section we apply the system of equations derived above and numerically study the Schwinger pair creation in the kinetic coupling model.

If the generated electromagnetic field and charged particles do not backreact on the universe expansion, the inflaton evolution can be determined independently. This simplified treatment is valid if the energy 
density of the produced electric field and charged particles is less than $\epsilon_{V}\rho_{\rm inf}$ \cite{Sobol:2018}, where $\epsilon_{V}=(M_{p}^{2}/2)(V'/V)^{2}$ is the slow-roll parameter and $\rho_{\rm inf}\simeq V(\phi)$ 
is the energy density of the inflaton. For the Starobinsky potential (\ref{Starobinsky-pot}), we get
\begin{equation}
\label{condition-BR}
\frac{\exp(2\beta\phi/M_{p})}{2e^{2}}\mathcal{E}^{2}+\rho_{\chi}\lesssim \mu^{2}M_{p}^{2} \exp\left(-2\sqrt{\frac{2}{3}}\frac{\phi}{M_{p}}\right).
\end{equation} 
If this condition is violated, the inflaton evolution cannot be treated separately and the backreaction has to be taken into account.

\section{Numerical results}
\label{sec-numerical}

\begin{figure}[h!]
	\centering
	\includegraphics[width=0.33\textheight]{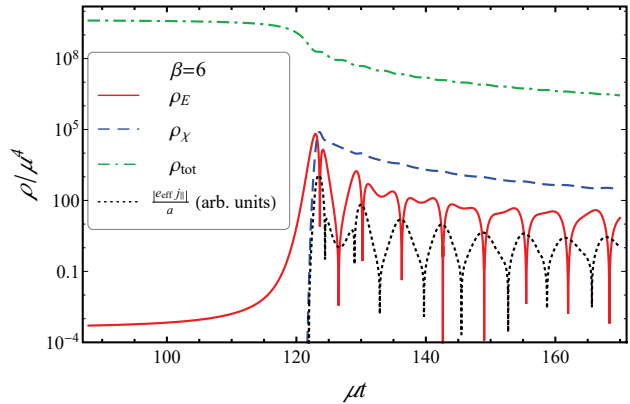}
	\caption{Time dependence of the electric energy density (red solid line), the energy density of charged scalar particles produced due to the Schwinger effect (blue dashed line), and the total energy density of the 
	universe (green dashed-dotted line) calculated in the kinetic approach for the coupling parameter $\beta=6$. Black dotted line shows the absolute value of the electric current (in arbitrary units). \label{fig-kinetic-beta6}}
\end{figure}

In this section, we present numerical solutions to the system of equations derived above. First, we consider a small value of the coupling parameter $\beta=6$ for which condition (\ref{condition-BR}) is satisfied 
during the whole period of inflation and after it, i.e., the backreaction does not occur. The time dependence of different components of the energy density is shown in Fig.~\ref{fig-kinetic-beta6}. The electric 
energy density is depicted by the red solid line, and the energy density of charged scalar particles produced due to the Schwinger effect is shown by the blue dashed line. The total energy density of the universe is given by the green 
dashed-dotted line and is, in fact, several orders of magnitude larger than the former ones.

As usual in the Ratra model, the electric energy density grows most intensively at the end of inflation (let us remind the reader that there are $N=60$ e-foldings before the end of inflation which ends at
$\mu t_{e}\approx 2N= 120$). The effective charge $e_{\rm eff}=e/f$ is suppressed by the Ratra coupling function unless the inflaton field is close to zero. Therefore, the Schwinger effect turns on only when 
inflation ends. The energy density of charged particles quickly grows until it reaches a value comparable to the energy density of the electric field. Produced charged particles, accelerated by a strong electric field, give 
rise to a current density (shown by the black dotted line in Fig.~\ref{fig-kinetic-beta6}) which affects the electric field evolution and the latter starts to decrease. It quickly reaches the zero value and changes direction; however, charged particles continue to move 
inertially in the same direction. Therefore, a retardation between the electric field and current appears which is clearly seen from Fig.~\ref{fig-kinetic-beta6}.  As a result, an oscillatory behavior of electric field emerges. This 
qualitatively new result was first discovered by us in Ref.~\cite{Gorbar:2019}.

It is interesting to compare our results with those in the literature. First of all, it was found~\cite{Kobayashi:2014,Hayashinaka:2016a} that the Schwinger current of scalar particles in a strong constant 
electric field $|\mathcal{E}|\gg H^{2}$ in de Sitter spacetime has the form
\begin{equation}
\label{j-Ohmic}
j_{\parallel}=-a e_{\rm eff}\frac{\mathcal{E}^{2}}{12\pi^{3}H} {\rm sign}(\mathcal{E}).
\end{equation}
It is possible to assume that the same functional dependence is true for a time-varying electric field and the electric current is determined by the electric field at the same moment of time. Obviously, the
electric current in this approach, which is widely used in the literature~\cite{Kobayashi:2014,Stahl:2018,Sobol:2018}, has a Markovian character. The electric field in this case is a 
decreasing function without sign-changing oscillations.

In Ref.~\cite{Gorbar:2019}, we attempted a kinetic description of the Schwinger effect in inflationary magnetogenesis. The Schwinger pair production was incorporated in the kinetic equation by means of a
local in time source term. More precisely, the right-hand side of Eq.~(\ref{eq-F-2}) was replaced with the following function of electric field and kinetic function $\mathcal{F}$:
\begin{equation}
\label{source}
\mathcal{S}(\mathcal{E},\mathcal{F}) =(1+2\mathcal{F})\sqrt{|\mathcal{E}|} \exp\left(-\pi\frac{m^{2}+\mathbf{p}^{2}}{|\mathcal{E}|}\right).
\end{equation}
This function was chosen requiring that it gives at $\mathcal{F}=0$ the correct expression for the pair production rate $\Gamma=\mathcal{E}^{2}\exp(-\pi m^{2}/|\mathcal{E}|)/(2\pi)^{3}$ \cite{Schwinger:1951,Nikishov:1970,Cohen:2008} and takes into 
account the Bose enhancement through the factor $(1+2\mathcal{F})$. 

Moreover, we considered~\cite{Gorbar:2019} also a hydrodynamical approach based on the following system of equations governing the evolution of the number density $n_{\chi}$, 
energy density $\rho_{\chi}$, and conduction current $j_{\rm cond}$ of produced particles:
\begin{eqnarray}
\frac{dn_{\chi}}{dt}+3Hn_{\chi}&=&2\Gamma,\label{n-eq}\\
\frac{d\rho_{\chi}}{dt}+4H\rho_{\chi}&=&\mathcal{E}(j_{\rm cond}+j_{\rm pol}),\label{rho-eq}\\
\frac{dj_{\rm cond}}{dt}+3Hj_{\rm cond}&=&\mathcal{E}\frac{n_{\chi}^{2}-j_{\rm cond}^{2}}{\rho_{\chi}}.\label{j-cond-eq}
\end{eqnarray}
Here the currents are normalized in such a way that the full current appearing on the right-hand side of Eq.~(\ref{eq-for-E}) is expressed as $j_{\parallel}=-ae_{\rm eff}(j_{\rm cond}+j_{\rm pol})$. The polarization current 
$j_{\rm pol}$ is caused by virtual particles and antiparticles in the process of their creation from vacuum. It can be estimated as $j_{\rm pol}\simeq {\rm sign}(\mathcal{E}) \Gamma/\sqrt{|\mathcal{E}|}$ with a numerical 
prefactor of order unity which depends on the specific form of the source function $\mathcal{S}(\mathcal{E},\mathcal{F})$. The hydrodynamical approach describes the retardation of the current with 
respect to the electric field; however, it does not take into account the effects of quantum statistics (i.e., the induced pair production for bosons and the Pauli blocking for fermions).

\begin{figure}[h!]
	\centering
	\includegraphics[width=0.33\textheight]{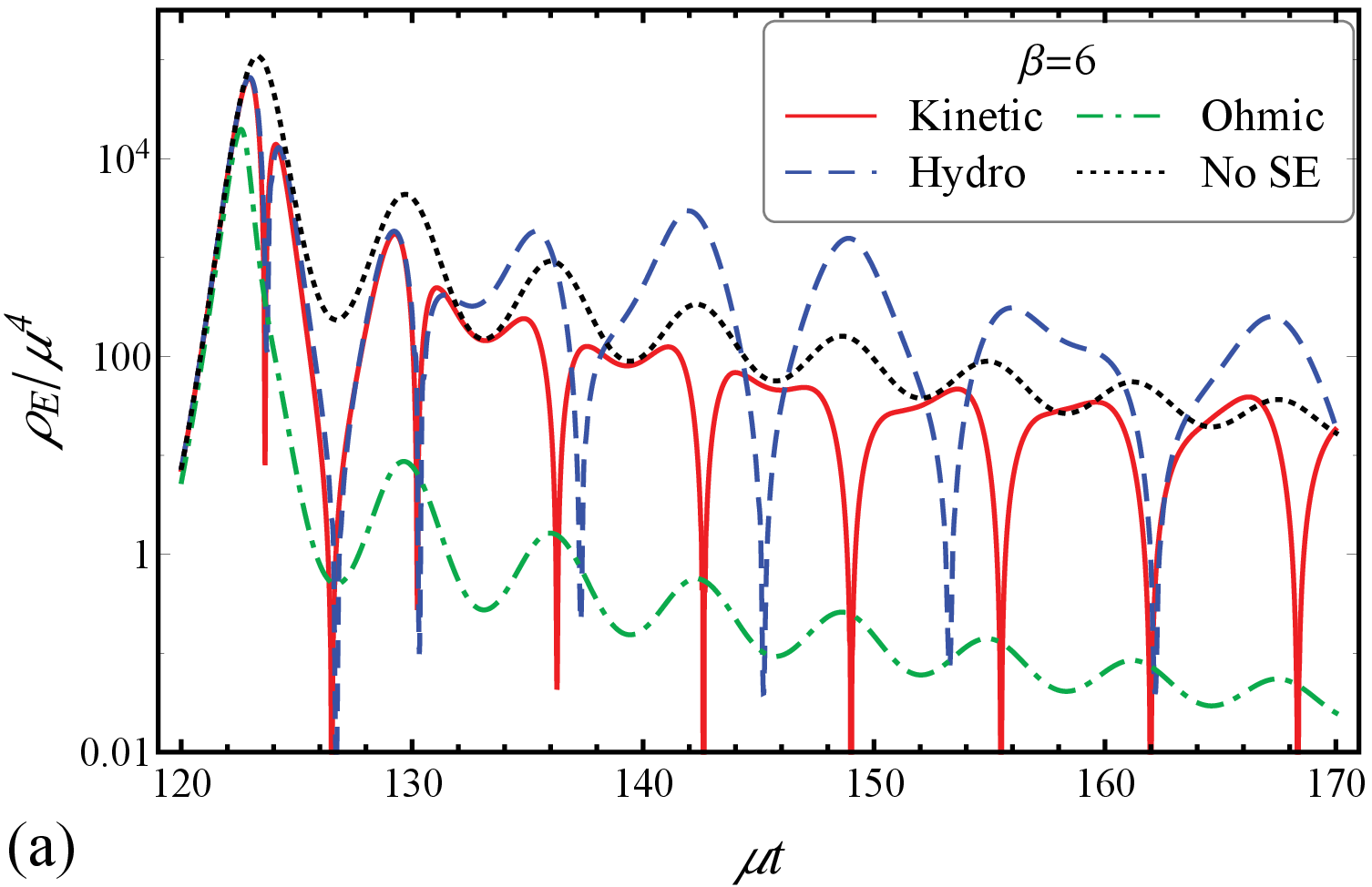}\hspace*{0.5cm}
	\includegraphics[width=0.33\textheight]{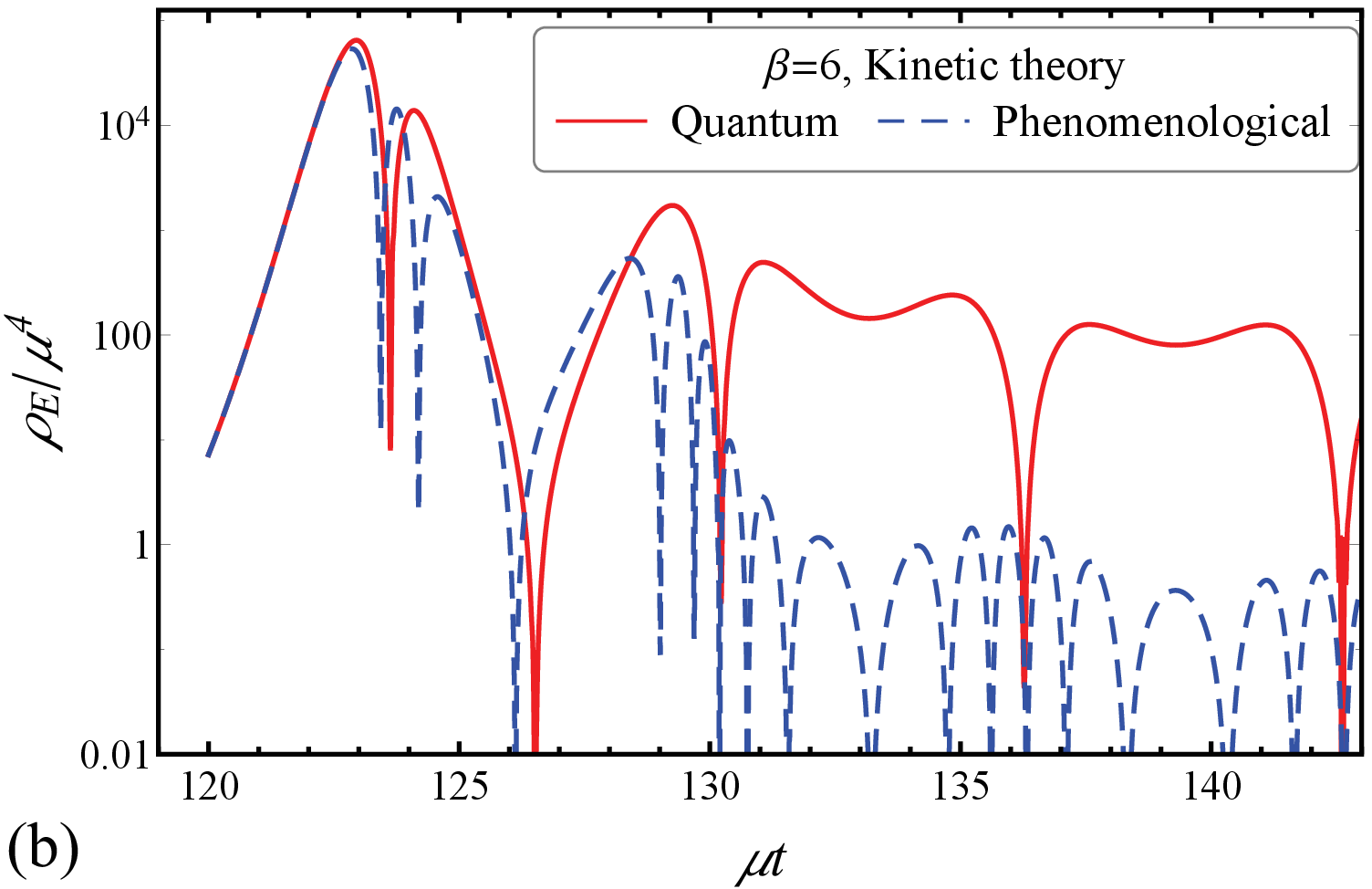}
	\caption{(a) Time dependence of the electric energy density calculated in the quantum kinetic theory (red solid line), in the hydrodynamical approach
	(blue dashed line), and with the 
	Schwinger current of the Ohmic form (\ref{j-Ohmic}) (green dashed-dotted line). Black dotted line shows the time dependence of the electric energy density in the absence of the Schwinger effect. (b) The
	evolution of the electric energy density in the quantum kinetic theory developed in this paper (red solid line) and in the phenomenological kinetic theory of Ref.~\cite{Gorbar:2019} (blue dashed line). 
	In both panels, the coupling parameter equals $\beta=6$.  \label{fig-compare-methods-beta6}}
\end{figure}

We compare the time evolution of the electric energy density calculated in all above-mentioned approaches with that in the first-principles quantum kinetic approach developed in this paper and shown 
by the red solid line in
Fig.~\ref{fig-compare-methods-beta6}. We compare it in panel (a) with the result of the hydrodynamical approach (blue dashed line) and with the case where the Schwinger current is given by Eq.~(\ref{j-Ohmic}) 
(green dashed-dotted line). The time evolution in the absence of the Schwinger effect is plotted by black dotted line. There are several features which we would like to point out. First of all, as we have already 
mentioned, the non-Markovian dependence of an electric current on the electric field leads to an oscillatory behavior in contrast to the case with the Ohmic current. Second, the frequency of oscillations of the electric 
energy density in the quantum kinetic theory is somewhat higher than in the hydrodynamical approach. This is because the effects of the quantum statistics are not taken into account 
in the latter case. 
In fact, the Bose enhancement of the Schwinger source which is encoded in the factor $(1+2\mathcal{F})$ on the right-hand side of Eq.~(\ref{eq-G-2}) leads to a faster growth of current and, as a result, to a more
rapid change of electric field. For the same reason, the amplitude of oscillations is lower in the quantum kinetic theory approach. Note that the results of both approaches are in 
good agreement during the first three oscillations when the kinetic function $\mathcal{F}$ is small compared to unity and the Bose enhancement is negligible.

Figure~\ref{fig-compare-methods-beta6}(b) compares the results obtained in the quantum kinetic approach developed in this paper (red solid line) and in the phenomenological kinetic theory~\cite{Gorbar:2019} (blue dashed 
line). Although they both predict an oscillatory behavior of the electric energy density, the frequency and amplitude of these oscillations drastically differ in these two cases. During the first few oscillations they 
more or less agree; however, later the phenomenological kinetic theory gives a rapidly oscillating solution with abruptly decreasing amplitude. This difference is caused by a different form of the Schwinger source. Indeed, the source term (\ref{source}) is local in time and momentum 
in the 
phenomenological kinetic theory~\cite{Gorbar:2019}. This means that the production of particles in a certain mode $\mathbf{p}$ is proportional to the 
filling factor $(1+2\mathcal{F}(\mathbf{p}))$ of this mode. Thus, if there are many particles with a given momentum, their production is strongly enhanced. In particular, 
this leads to continuous and avalanchelike enhancement
of modes with small momenta $p\lesssim \sqrt{|\mathcal{E}|}$ leading to a high and sharp peak in the particle distribution. In contrast to this, in the quantum kinetic 
approach of the present
paper, we have a source term which is nonlocal both in time and momentum. Although the Bose enhancement factor is not present directly on the right-hand side of the quantum Vlasov equation (\ref{eq-F-2}), it is hidden in another kinetic equation
describing the evolution of the source term. As a result, the particle production is determined by the distribution function in all preceding moments of time and for all possible momenta. This smooths peaks in 
the distribution function and leads to slower changes of current. 

We would like to remind the reader that the hydrodynamical approach (\ref{n-eq})--(\ref{j-cond-eq}) was derived from the phenomenological kinetic theory with the local source term (\ref{source}) adopting 
some additional approximations: (i) neglecting the Bose enhancement and (ii) assuming that all particles have the same momentum. Nevertheless, for small $\beta$, we have a paradoxical 
situation when the result of the phenomenological kinetic theory 
is in worse accordance with the first-principles quantum kinetic theory prediction than the result of the hydrodynamical approach. This can be explained by the fact that, for small $\beta$, the avalanchelike production of the particles caused by the local factor $(1+2\mathcal{F})$ in the phenomenological kinetic theory does not occur in the real situation. In fact, for $\beta=6$, the maximal value of the enhancement factor is of order 10 and is achieved only for an extremely small range of momenta. Typically, however, $(1+2\mathcal{F})= \mathcal{O}(1)$ and the hydrodynamical description appears to be quite accurate. Naturally, this approximation fails for larger values of $\beta$.

\begin{figure}[h!]
	\centering
	\includegraphics[width=0.33\textheight]{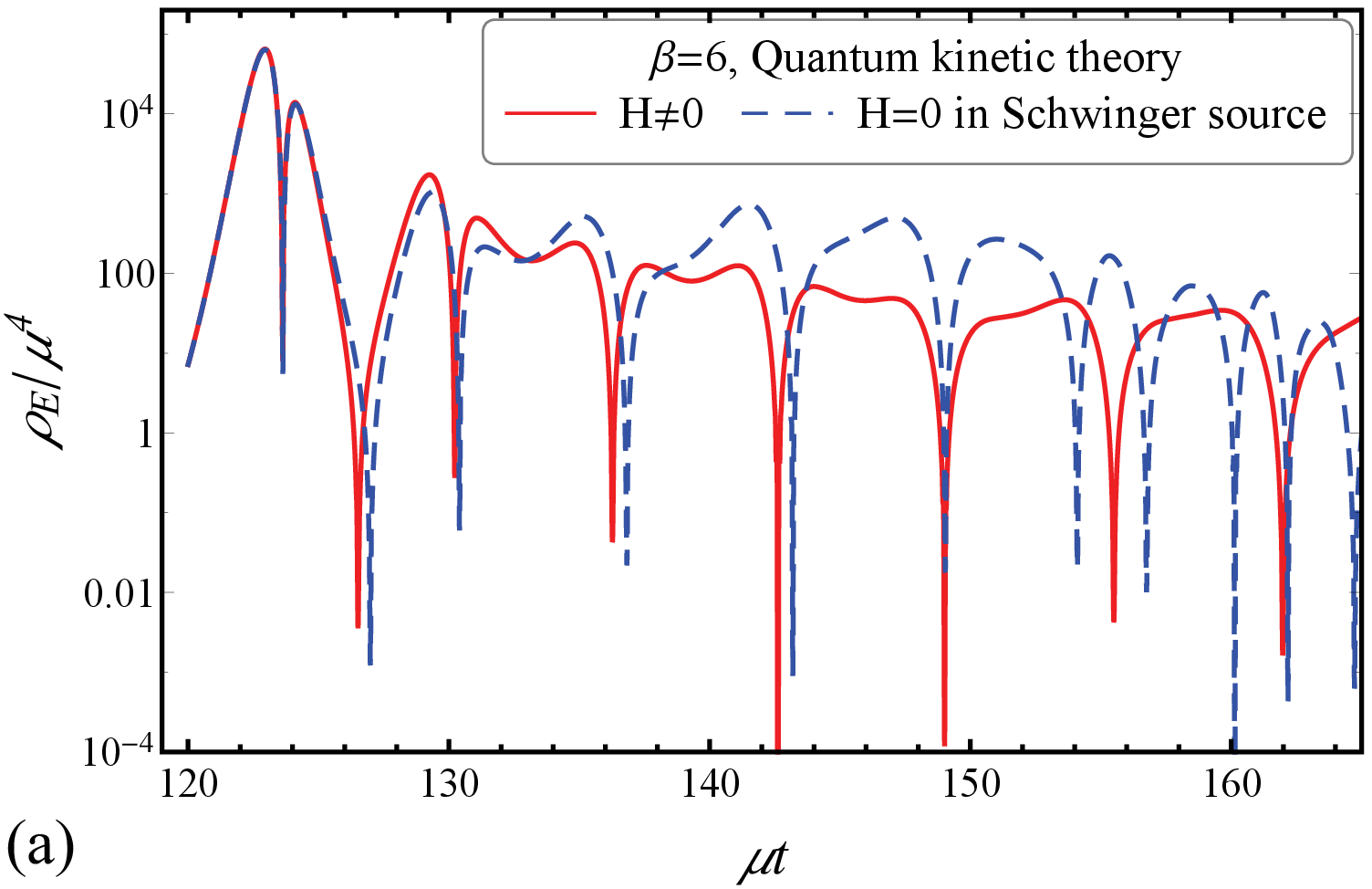}\hspace*{0.5cm}
	\includegraphics[width=0.33\textheight]{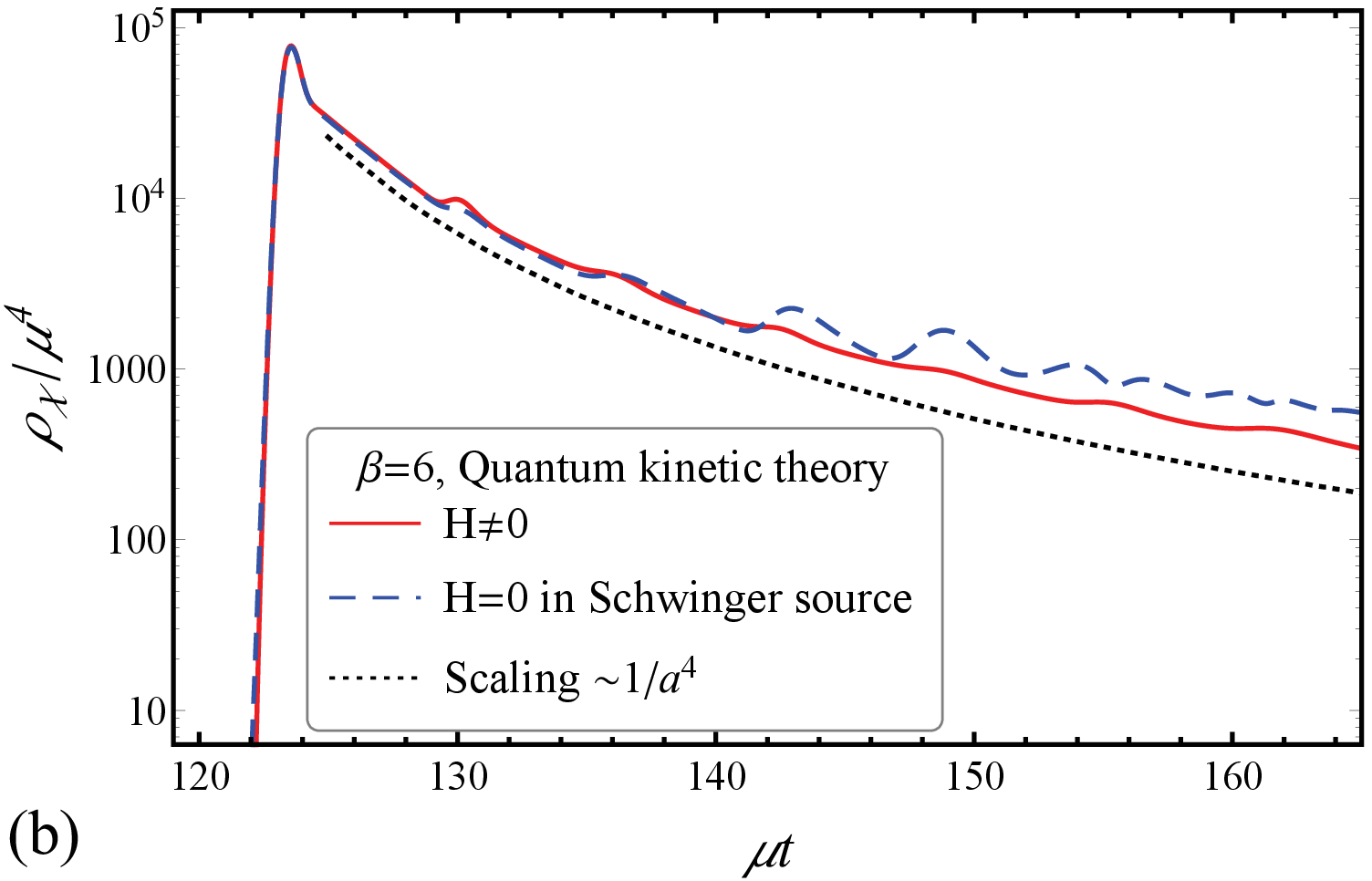}
	\caption{Time evolution of the electric energy density (a) and the energy density of scalar particles produced due to the Schwinger effect (b) calculated in the quantum kinetic theory described in an expanding universe (red solid line) and in the case when the expansion is neglected in the Schwinger source (blue dashed line). Black dotted line in (b) shows the scaling $\sim 1/a^{4}(t)$ in an expanding universe. The coupling parameter equals $\beta=6$. \label{fig-compare-no-expansion-beta6}}
\end{figure}

It is often assumed in the literature that the universe expansion can be neglected in the description of the Schwinger pair production process. For instance, in the phenomenological kinetic theory of Ref.~\cite{Gorbar:2019}, the Schwinger source term in the kinetic equation was taken in the same form as in flat spacetime. 
In the quantum kinetic theory of this work the analog of the Schwinger source is the right-hand side of Eq.~(\ref{eq-F-2}) which contains the product $Q(t,\mathbf{p})\mathcal{G}(t,\mathbf{p})$. Neglecting the universe expansion in this term means setting $H=0$ in the definition of $Q$-function (\ref{Q-1/6}) as well as in Eqs.~(\ref{eq-G-2})--(\ref{eq-H-2}).
To check the validity of this assumption, we determined the electric energy density and the energy density of scalar particles produced by the Schwinger effect in the approximation described above. These are shown by blue dashed lines in Fig.~\ref{fig-compare-no-expansion-beta6}(a) and \ref{fig-compare-no-expansion-beta6}(b), respectively, in comparison with the case of full quantum kinetic theory in an expanding universe (red solid lines).
During the first few oscillations (when the electric field is the strongest) the results are in good agreement, while later the particle production is slightly more effective in the case when the expansion is neglected. This can be explained by the following arguments. In the full quantum kinetic theory the Schwinger source is nonlocal in time and momentum because it takes into account that the virtual particles and antiparticles are created at some earlier moments of time. Before becoming real they are accelerated by the electric field and their momenta are redshifted due to the universe expansion. If the latter effect is neglected, the energy density of newly produced particles is greater than in the case with expansion. However, this effect is important only if the pair creation time is sufficiently large so that this redshift is significant, i.e. in the weak electric field. In fact, the deviation between the dashed and solid curves in Fig.~\ref{fig-compare-no-expansion-beta6} (b) is significant only at late times when the electric field is weak.

\begin{figure}[h!]
	\centering
	\includegraphics[width=0.33\textheight]{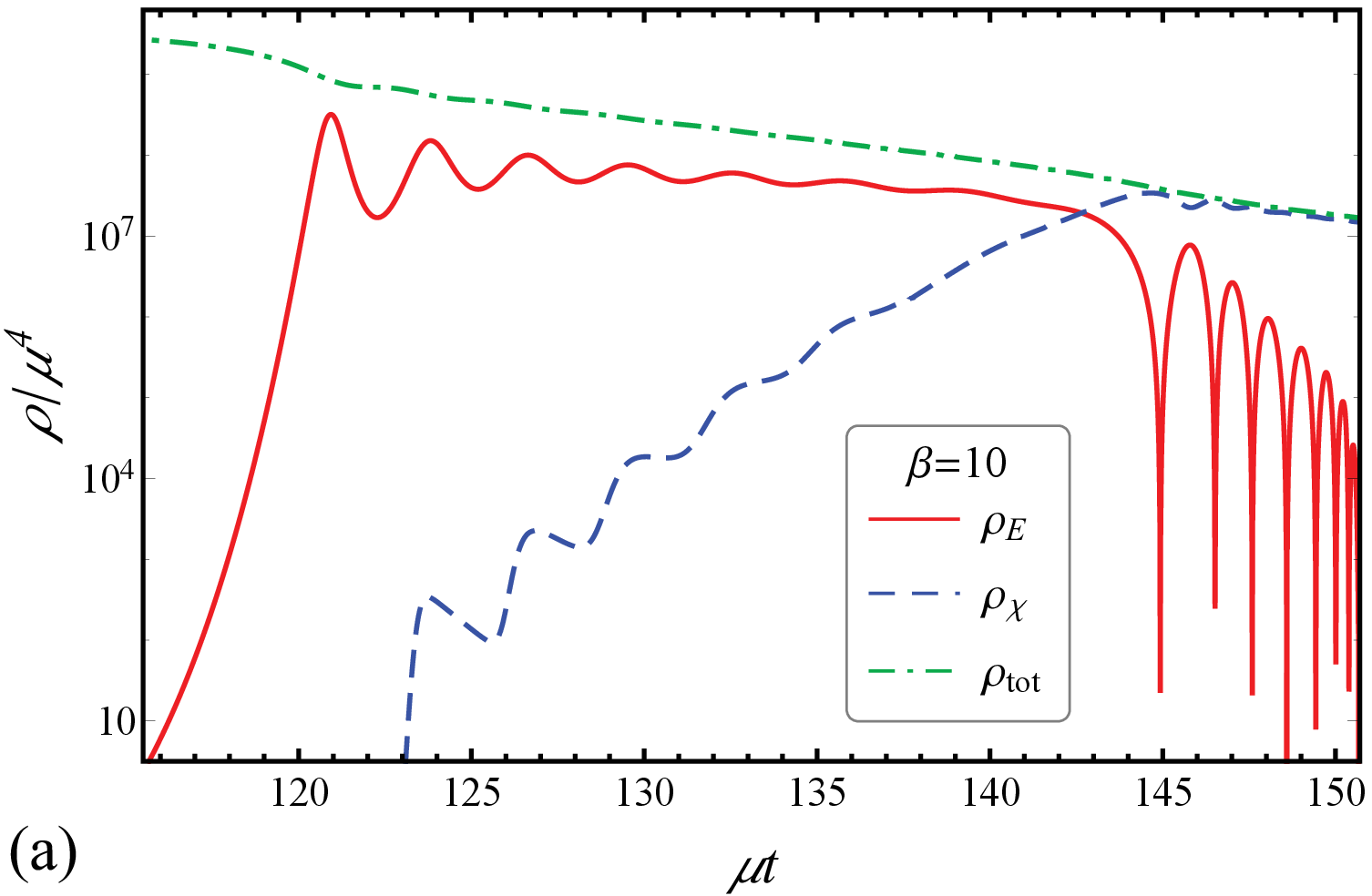}\hspace*{0.5cm}
	\includegraphics[width=0.33\textheight]{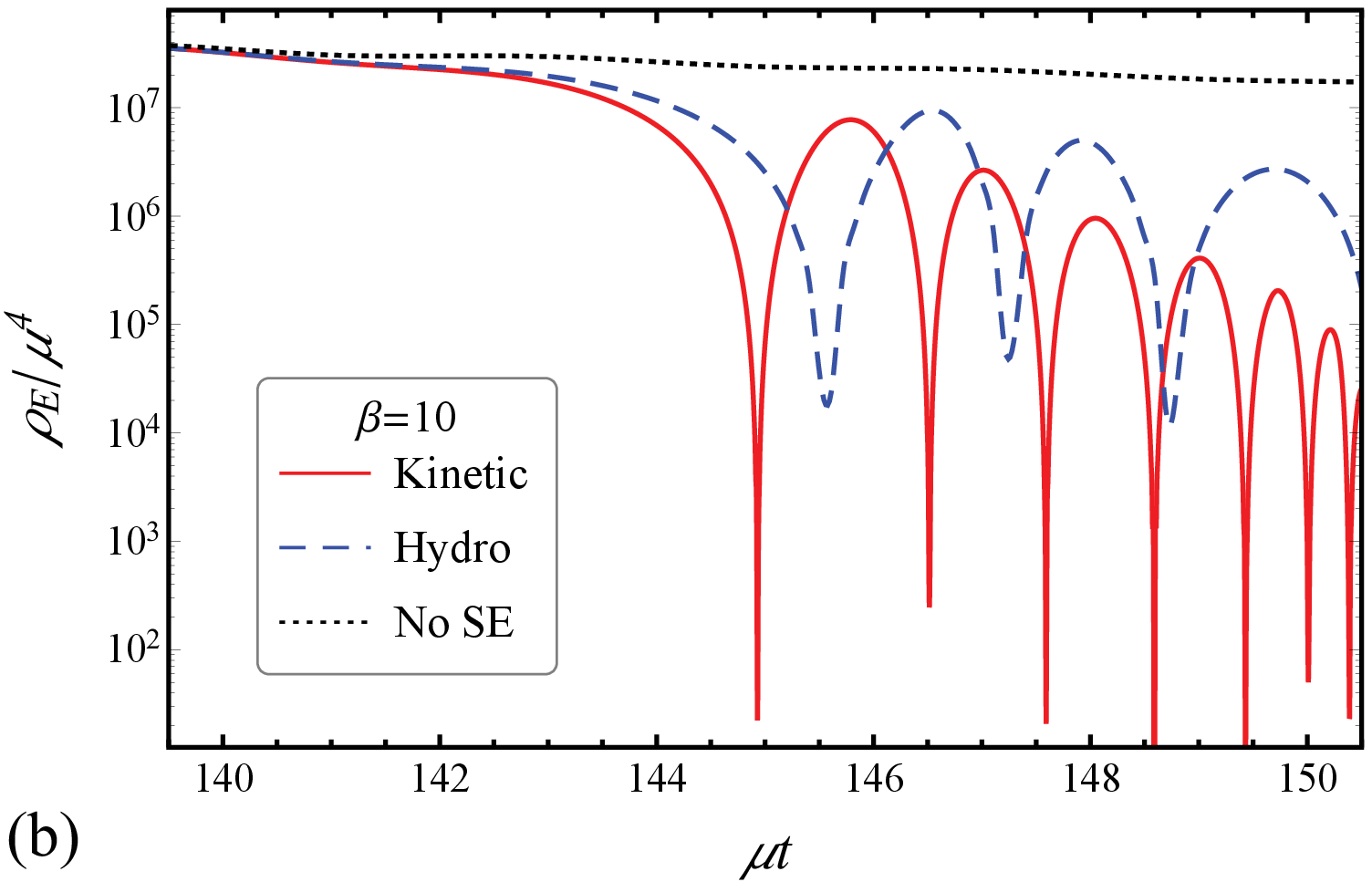}
	\caption{(a) Time dependence of the electric energy density (red solid line), the energy density of scalar charged particles produced due to the Schwinger effect (blue dashed line), and the total energy density of 
	the universe (green dashed-dotted line) calculated in the kinetic approach. (b) Time dependence of the electric energy density calculated in the quantum kinetic method (red
	solid line) and the 
	hydrodynamical approach (blue dashed line). Black dotted line shows the time dependence of the electric energy density in the absence of the Schwinger effect. In both panels, the coupling parameter $\beta=10$.  
	\label{fig-beta10}}
\end{figure}

Finally, let us say a few words about the backreaction. For the coupling parameter $\beta=10$, we plot the time dependence of the electric energy density (red solid line), the energy density of charged particles 
(blue dashed line), and the total energy density (green dashed-dotted line) in Fig.~\ref{fig-beta10}(a). One can see that at $\mu t\approx 120$ the electric energy density becomes close to that of the inflaton and the 
backreaction regime occurs. In this regime the energy density of the universe is dominated first by electric field and later by charged particles produced due to the Schwinger effect realizing the Schwinger 
reheating scenario. The electric field demonstrates an oscillatory behavior, but with the frequency of oscillations much higher than for $\beta=6$. This is because the electric field is much stronger now and
produces scalar particles much more effectively. It was shown~\cite{Gorbar:2019} that the period of oscillations equals $t_{\rm osc}\sim (e_{\rm eff}T)^{-1}$, where temperature $T\sim \rho_{\chi}^{1/4}$ is 
determined by the energy density of produced particles. Since the latter is 4 orders of magnitude larger for $\beta=10$ than for $\beta=6$, the oscillation frequency is almost 10 times higher in the former case.

Figure~\ref{fig-beta10}(b) compares the time dependence of the electric energy density calculated in the first-principles quantum kinetic method (red solid line) with that in the hydrodynamical 
approach (blue dashed line). As in the case $\beta=6$, here we also observe that the frequency of oscillations in the hydrodynamical approach are less 
frequent because the Bose enhancement is not taken into account.

\section{Conclusion}
\label{sec-concl}

Using the first-principles quantum kinetic method, we studied the dynamics of the Schwinger effect for charged scalar particles in an inflation-produced electric field. Calculating the Bogolyubov 
coefficients, we derived a set of quantum
kinetic equations describing the process of the pair creation. The quantum Vlasov equation (\ref{eq-F-2}) for the momentum distribution function $\mathcal{F}$ acquires a source term which is nonlocal in time 
and determined by Eqs.~(\ref{eq-G-2})--(\ref{eq-H-2}). This system takes into account the universe expansion and the effects of quantum statistics (Bose enhancement).

We investigated the asymptotic behavior of the kinetic functions at large momentum
and determined UV divergences in the electric current and the energy-momentum tensor. Using the dimensional regularization, we found
counterterms ensuring the finiteness of the physical quantities
which enter the Maxwell and Friedmann equations. The latter together with the three kinetic equations describe self-consistently the evolution of the electric field and produced charged scalar particles in an expanding universe. 
This system of equations is the central result of our paper.

There are several approaches to the description of the Schwinger effect in a homogeneous electric field in the early universe with a different amount of assumptions. The present work finalizes their hierarchy:
\begin{enumerate}
	\item The most straightforward and naive approach is to use the Schwinger current calculated in a constant electric field in de Sitter spacetime \cite{Kobayashi:2014,Hayashinaka:2016a,Hayashinaka:2016b}. 
	In the strong field regime $eE\gg H^{2}$, it is given by Eq.~(\ref{j-Ohmic}). Generalizing this dependence to the case of a time-varying electric field, one gets a local in time Schwinger current.  
	\item The hydrodynamical approach, based on Eqs.~(\ref{n-eq})--(\ref{j-cond-eq}), was proposed in Ref.~\cite{Gorbar:2019}. Here, the conduction current is nonlocal in time and is determined from an ordinary differential equation coupled to the Maxwell equation. This implies a qualitatively new effect of retardation of the current with respect to electric field. This approximation, however, does not take into account the effects of quantum statistics on the pair creation process.
	\item The phenomenological kinetic theory considered in Ref.~\cite{Gorbar:2019} also leads to a retarded conduction current which is determined from the Boltzmann kinetic equation with the source term local in time (\ref{source}). The source was constructed phenomenologically taking into account the quantum statistics of created particles and assuming that the creation rate depends only on the filling number of a given mode at the same moment of time. 
	\item Finally, in the present work, the source term on the right-hand side of Eq.~(\ref{eq-F-2}) is nonlocal in time itself and is determined from two additional kinetic equations (\ref{eq-G-2})--(\ref{eq-H-2}). Being derived from the quantum field theory, the Schwinger source now takes into account the non-Markovian character of the pair creation process. Moreover, it captures the impact of the universe expansion even on the momenta of a \textit{virtual} particle and antiparticle being accelerated by the electric field. Although this effect is small in the strong electric field $|e\mathbf{E}| \gg H^2$, it may significantly decrease the energy density of charged particles produced by a weak electric field.
\end{enumerate}

As a test bed for our formalism, we used the kinetic coupling model $\mathcal{L}_{\rm EM}=-f^{2}(\phi)F_{\mu\nu}F^{\mu\nu}/4$ of the electromagnetic field to the inflaton because it naturally provides a strong 
long-range electric field which then drives the Schwinger effect. In fact, due to decreasing in time coupling function $f$, electric fields generated in this model are much stronger than the magnetic ones so that the 
latter can be neglected \cite{Kanno:2009,Demozzi:2009,Sobol:2018}. We use the one-parametric Ratra coupling function $f(\phi)=\exp(\beta\phi/M_{p})$ which can produce electric fields of different strengths. 

For small coupling parameter $\beta=6$, the electric energy density does not cause the backreaction on the background evolution. In this case, the energy density of created charged particles is also much less than that of 
the inflaton and the Schwinger effect plays a small role in the universe reheating. In accordance with our previous study \cite{Gorbar:2019}, we observe an oscillatory behavior of the electric field which can be explained 
by the inertial properties of the electric current. During these oscillations the amplitude of the electric field slowly decreases due to the universe expansion in contrast to a naive expectation that the Schwinger effect 
should abruptly reduce the electric field. These are good news from the viewpoint of magnetogenesis because the magnetic field can in principle be enhanced by a (much stronger) oscillating electric field 
\cite{Kobayashi:2019}.

For larger coupling parameter $\beta=10$, the electric field is much stronger and affects the universe expansion slowing down the inflaton and prolongating the inflation stage. Moreover, charged particles 
produced by this strong electric field constitute a large fraction of the energy density of the universe. This leads to the Schwinger reheating \cite{Tangarife:2017,Sobol:2018,Gorbar:2019} which must be also 
considered as a complementary scenario in addition to the usual ones with the inflaton decay and parametric resonance \cite{Kofman:1997,Amin:2015}. The electric field also exhibits 
an oscillatory behavior; however, the frequency is much higher because of the higher particle production rate.

In our work, we considered by using the quantum kinetic method only the case of scalar charged particles produced by the electric field. Needless to say, an extension to the 
case of fermionic charge carriers is definitely the next necessary step. In addition, it would be important to take into account the thermalization of produced particles due to collisions and the reduction of the electric 
current due to annihilation processes of the charge carriers. These issues are technically complicated and deserve a separate investigation. Another important issue is the axial coupling model
$\mathcal{L}_{\rm axial}=-I(\phi) F_{\mu\nu}\tilde{F}^{\mu\nu}$, where generated magnetic fields can be as strong as 
the electric ones \cite{Garretson:1992,Durrer:2011,Domcke:2018,Domcke:2019}. Therefore, it is much desirable and would be very interesting to study the 
impact of the magnetic field on the Schwinger pair creation in the first-principles quantum kinetic approach. We plan to address this issue elsewhere.

\begin{acknowledgments}
	
	We are grateful to Yu.V.~Shtanov for valuable comments.
	The work of O.~O.~S. was supported by the Swiss National Science Foundation Grant No. 200020B\_182864.	
	The work of S.~I.~V. was supported by the Swiss National Science Foundation Grant No. SCOPE IZSEZ0-186551 and by the German Academic Exchange Service (DAAD), Grant No. 57387479. 
	S.~I.~V. is grateful to Professor Mikhail Shaposhnikov for his kind hospitality at the Institute of Physics, \'{E}cole Polytechnique F\'{e}d\'{e}rale de Lausanne, Switzerland, where the part of this work was done.

\end{acknowledgments}

\appendix

\section{Asymptotic UV expansions of kinetic functions}
\label{app-UV}

In this appendix, we investigate the asymptotic UV behavior of the kinetic functions satisfying the system of Eqs.~(\ref{eq-F-2})--(\ref{eq-H-2}). First of all, we expand all coefficients in the kinetic equations 
in inverse powers of momentum, i.e.,
\begin{equation}
\label{eps-expansion}
\omega(t,\mathbf{p})=p+\omega^{(-1)}+\mathcal{O}(p^{-3}), 
\end{equation}
\begin{equation}
\label{Q-expasion}
Q(t,\mathbf{p})=Q^{(0)}+Q^{(-1)}+Q^{(-2)}+\mathcal{O}(p^{-3}),
\end{equation}
where
\begin{eqnarray}
\omega^{(-1)}&=&\frac{4m^{2}+(24\xi-6)\dot{H}+(48\xi-9)H^{2}}{8p}, \label{eps-1}\\
Q^{(0)}&=&-H, \label{Q-0}\\
Q^{(-1)}&=&\frac{\boldsymbol{\mathcal{E}}\cdot\mathbf{p}}{p^{2}},\label{Q-1}\\
Q^{(-2)}&=&\frac{4Hm^{2}+(12\xi-3)\ddot{H}+(72\xi-15)H\dot{H}+(48\xi-9)H^{3}}{4p^{2}}.\label{Q-2}
\end{eqnarray}
We decompose also the complete time derivative on the left-hand side of Eqs.~(\ref{eq-F-2})--(\ref{eq-H-2}) into two differential operators
\begin{equation}
\frac{d}{dt}=\hat{L}^{(0)}+\hat{L}^{(-1)}, \quad \hat{L}^{(0)}=\frac{\partial}{\partial t}-H\mathbf{p}\frac{\partial}{\partial \mathbf{p}},\quad \hat{L}^{(-1)}=\boldsymbol{\mathcal{E}}\frac{\partial}{\partial \mathbf{p}}.
\end{equation}
Obviously, the application of the operator $\hat{L}^{(0)}$ does not change the UV behavior of the corresponding functions while the application of $\hat{L}^{(-1)}$ decreases the power of momentum by unity.

In the beginning of Sec.~\ref{subsec-UV}, we determined the leading terms of the power expansion of the kinetic functions. We have
\begin{eqnarray}
\mathcal{F}(t,\mathbf{p})&=&\mathcal{F}^{(-2)}+\mathcal{F}^{(-3)}+\mathcal{F}^{(-4)}+\mathcal{O}(p^{-5}),\label{dec-f}\\
\mathcal{G}(t,\mathbf{p})&=&\mathcal{G}^{(-2)}+\mathcal{G}^{(-3)}+\mathcal{G}^{(-4)}+\mathcal{O}(p^{-5}),\label{dec-g}\\
\mathcal{H}(t,\mathbf{p})&=&\mathcal{H}^{(-1)}+\mathcal{H}^{(-2)}+\mathcal{H}^{(-3)}+\mathcal{O}(p^{-4}).\label{dec-h}
\end{eqnarray}
Substituting these expansions together with Eqs.~(\ref{eps-expansion})--(\ref{Q-2}) into the Vlasov equations (\ref{eq-F-2})--(\ref{eq-H-2}) and separating terms with different powers of momentum, we obtain the 
following chain of equations:
\begin{eqnarray}
&&2p\mathcal{H}^{(-1)}+\frac{1}{2}Q^{(0)}=0,\label{eq-h-1}\\
&&2p\mathcal{H}^{(-2)}+\frac{1}{2}Q^{(-1)}=0, \label{eq-h-2}\\
&&2p\mathcal{H}^{(-3)}+2\omega^{(-1)}\mathcal{H}^{(-1)}+Q^{(0)}\mathcal{F}^{(-2)}+\frac{1}{2}Q^{(-2)}-\hat{L}^{(0)}\mathcal{G}^{(-2)}=0,\label{eq-h-3}\\
&&\hat{L}^{(0)}\mathcal{H}^{(-1)}+2p\mathcal{G}^{(-2)}=0,\label{eq-g-2}\\
&&\hat{L}^{(0)}\mathcal{H}^{(-2)}+\hat{L}^{(-1)}\mathcal{H}^{(-1)}+2p\mathcal{G}^{(-3)}=0,\label{eq-g-3}\\
&&\hat{L}^{(0)}\mathcal{H}^{(-3)}+\hat{L}^{(-1)}\mathcal{H}^{(-2)}+2p\mathcal{G}^{(-4)}+2\omega^{(-1)}\mathcal{G}^{(-2)}=0,\label{eq-g-4}\\
&&\hat{L}^{(0)}\mathcal{F}^{(-2)}-Q^{(0)}\mathcal{G}^{(-2)}=0,\label{eq-f-2}\\
&&\hat{L}^{(0)}\mathcal{F}^{(-3)}+\hat{L}^{(-1)}\mathcal{F}^{(-2)}-Q^{(0)}\mathcal{G}^{(-3)}-Q^{(-1)}\mathcal{G}^{(-2)}=0,\label{eq-f-3}\\
&&\hat{L}^{(0)}\mathcal{F}^{(-4)}+\hat{L}^{(-1)}\mathcal{F}^{(-3)}-Q^{(0)}\mathcal{G}^{(-4)}-Q^{(-1)}\mathcal{G}^{(-3)}-Q^{(-2)}\mathcal{G}^{(-2)}=0.\label{eq-f-4}
\end{eqnarray}

Equation~(\ref{eq-h-1}) gives
\begin{equation}
\label{h-1}
\mathcal{H}^{(-1)}=-\frac{1}{4p}Q^{(0)}=\frac{H}{4p}.
\end{equation}
While Eq.~(\ref{eq-h-2}) implies
\begin{equation}
\label{h-2}
\mathcal{H}^{(-2)}=-\frac{1}{4p}Q^{(-1)}=-\frac{\mathbf{p}\cdot\boldsymbol{\mathcal{E}}}{4p^{3}},
\end{equation}
Equation~(\ref{eq-g-2}) leads to
\begin{equation}
\label{g-2}
\mathcal{G}^{(-2)}=-\frac{1}{2p}\hat{L}^{(0)}\mathcal{H}^{(-1)}=-\frac{1}{2p}\left[\frac{\partial}{\partial t}-H\mathbf{p}\frac{\partial}{\partial \mathbf{p}}\right]\frac{H}{4p}=-\frac{\dot{H}+H^{2}}{8p^{2}}.
\end{equation}
Equation~(\ref{eq-g-3}) produces
\begin{eqnarray}
\mathcal{G}^{(-3)}&=&-\frac{1}{2p}\left[\hat{L}^{(0)}\mathcal{H}^{(-2)}+\hat{L}^{(-1)}\mathcal{H}^{(-1)}\right]=\nonumber\\
&=&\frac{1}{2p}\left[\frac{\partial}{\partial t}-H\mathbf{p}\frac{\partial}{\partial \mathbf{p}}\right]\frac{\mathbf{p}\cdot\boldsymbol{\mathcal{E}}}{4p^{3}}-\frac{1}{2p}\boldsymbol{\mathcal{E}}\frac{\partial}{\partial \mathbf{p}}\frac{H}{4p}=\frac{\mathbf{p}\cdot\big(\dot{\boldsymbol{\mathcal{E}}}+3H\boldsymbol{\mathcal{E}}\big)}{8p^{4}}.\label{g-3}
\end{eqnarray}
It follows from Eq.~(\ref{eq-f-2}) that
\begin{equation}
\label{f-2}
\hat{L}^{(0)}\mathcal{F}^{(-2)}=Q^{(0)}\mathcal{G}^{(-2)}=\frac{H}{2p}\hat{L}^{(0)}\frac{H}{4p}= \hat{L}^{(0)}\left(\frac{H}{4p}\right)^{2},\quad \Rightarrow\quad \mathcal{F}^{(-2)}=\frac{H^{2}}{16p^{2}}.
\end{equation}
Equation~(\ref{eq-f-3}) provides
\begin{eqnarray}
\hat{L}^{(0)}\mathcal{F}^{(-3)}&=&Q^{(0)}\mathcal{G}^{(-3)}+Q^{(-1)}\mathcal{G}^{(-2)}-\hat{L}^{(-1)}\mathcal{F}^{(-2)}=\nonumber\\
&=&-\frac{H}{2p}\hat{L}^{(0)}\frac{\mathbf{p}\cdot\boldsymbol{\mathcal{E}}}{4p^{3}}+\frac{H}{2p}\hat{L}^{(-1)}\frac{H}{4p}-\frac{\boldsymbol{\mathcal{E}}\cdot\mathbf{p}}{2p^{3}}\hat{L}^{(0)}\frac{H}{4p}-\hat{L}^{(-1)}\left(\frac{H}{4p}\right)^{2}=\nonumber\\
&=&-2\left[\frac{H}{4p}\hat{L}^{(0)}\frac{\mathbf{p}\cdot\boldsymbol{\mathcal{E}}}{4p^{3}}+\frac{\mathbf{p}\cdot\boldsymbol{\mathcal{E}}}{4p^{3}}\hat{L}^{(0)}\frac{H}{4p}\right]=\hat{L}^{(0)}\frac{-H\mathbf{p}\cdot\boldsymbol{\mathcal{E}}}{8p^{4}}, \quad \Rightarrow\quad \mathcal{F}^{(-3)}=-\frac{H\mathbf{p}\cdot\boldsymbol{\mathcal{E}}}{8p^{4}}.\label{f-3}
\end{eqnarray}
Equation~(\ref{eq-h-3}) implies
\begin{eqnarray}
\mathcal{H}^{(-3)}&=&\frac{1}{2p}\left[\hat{L}^{(0)}\mathcal{G}^{(-2)}-\frac{1}{2}Q^{(-2)}-Q^{(0)}\mathcal{F}^{(-2)}-2\omega^{(-1)}\mathcal{H}^{(-1)}\right]=\nonumber\\
&=&\frac{(1-6\xi)\big(\ddot{H}+7H\dot{H}+6H^{3}\big)-3m^{2}H}{8p^{3}}.\label{h-3}
\end{eqnarray}
Equation~(\ref{eq-g-4}) leads to
\begin{eqnarray}
\mathcal{G}^{(-4)}\!\!\!&=&\!\!\!-\frac{1}{2p}\left[\hat{L}^{(0)}\mathcal{H}^{(-3)}+\hat{L}^{(-1)}\mathcal{H}^{(-2)}+2\omega^{(-1)}\mathcal{G}^{(-2)}\right]=\frac{\boldsymbol{\mathcal{E}}^{2}-3(\mathbf{p}\cdot\boldsymbol{\mathcal{E}})^{2}/p^{2}}{8p^{4}}-\\
\!\!\!&-&\!\!\!(1-6\xi)\frac{\dddot{H}+8\dot{H}^{2}+10H\ddot{H}+42H^{2}\dot{H}+20H^{4}}{16p^{4}}-\frac{2\dot{H}^{2}+3H^{2}\dot{H}+H^{4}-40 m^{2}H^{2}-16m^{2}\dot{H}}{64p^{4}}.\nonumber
\end{eqnarray}
Finally, Eq.~(\ref{eq-f-4}) gives
\begin{eqnarray}
\hat{L}^{(0)}\mathcal{F}^{(-4)}&=&Q^{(0)}\mathcal{G}^{(-4)}+Q^{(-1)}\mathcal{G}^{(-3)}+Q^{(-2)}\mathcal{G}^{(-2)}-\hat{L}^{(-1)}\mathcal{F}^{(-3)}=\nonumber\\
&=&\frac{(\mathbf{p}\cdot\boldsymbol{\mathcal{E}})(\mathbf{p}\cdot\dot{\boldsymbol{\mathcal{E}}})}{8p^{6}}+\frac{H(\mathbf{p}\cdot\boldsymbol{\mathcal{E}})^{2}}{4p^{6}}+\nonumber\\
&+&\frac{4H\dddot{H}+64 H\dot{H}^{2}+46H^{2}\ddot{H}+219H^{3}\dot{H}+99H^{5}+6\dot{H}\ddot{H}-24m^{2}H\dot{H}-48m^{2}H^{3}}{64p^{4}}-\nonumber\\
&-&\frac{3\xi\big(H\dddot{H}+14 H\dot{H}^{2}+11H^{2}\ddot{H}+52H^{3}\dot{H}+24H^{5}+\dot{H}\ddot{H}\big)}{8p^{4}}.\label{eq-f-4-gen}
\end{eqnarray}
We look for a solution to Eq.~(\ref{eq-f-4-gen}) in the following form:
\begin{equation}
\label{f-4-sol}
\mathcal{F}^{(-4)}=\frac{(\mathbf{p}\cdot\boldsymbol{\mathcal{E}})^{2}}{16p^{6}}+\frac{\Psi}{64 p^{4}}, \qquad \Psi=c_{1}H\ddot{H}+c_{2}H^{2}\dot{H}+c_{3}\dot{H}^{2}+c_{4}H^{4}-c_{5}m^{2}H^{2}.
\end{equation}
Substituting Eq.~(\ref{f-4-sol}) into Eq.~(\ref{eq-f-4-gen}), we arrive at the system of 8 linear equations for 5 coefficients $c_{i}$
\begin{eqnarray}
c_{1}&=&4(1-6\xi),\label{A-30}\\
c_{1}+2c_{3}&=&6(1-4\xi),\\
2c_{2}+4c_{3}&=&16(4-21\xi),\\
4c_{1}+c_{2}&=&2(23-132\xi),\\
4c_{2}+4c_{4}&=&219-1248\xi,\\
4c_{4}&=&99-576\xi,\\
2c_{5}&=&24,\\
4c_{5}&=&48.\label{A-37}
\end{eqnarray}
Among Eqs.~(\ref{A-30})--(\ref{A-37}), only five are independent. Hence, the following unique solution for this system exists:
\begin{equation}
c_{1}=4(1-6\xi), \quad c_{2}=28(1-6\xi)+2, \quad c_{3}=1,\quad c_{4}=24(1-6\xi)+\frac{3}{4}, \quad c_{5}=12.
\end{equation}
Therefore, we obtain
\begin{equation}
\mathcal{F}^{(-4)}=\frac{(\mathbf{p}\cdot\boldsymbol{\mathcal{E}})^{2}}{16p^{6}}+(1-6\xi)\frac{H\ddot{H}+7H^{2}\dot{H}+6H^{4}}{16p^{4}}+\frac{\dot{H}^{2}+2H^{2}\dot{H}+\frac{3}{4}H^{4}-12 m^{2}H^{2}}{64p^{4}}.
\end{equation}

Thus, we found a few leading terms in the asymptotic behavior of the kinetic functions $\mathcal{F}$, $\mathcal{G}$, and $\mathcal{H}$ for $p\to\infty$. However, the decomposition in inverse 
powers of momentum is not very useful for the calculation of the electric current, energy density, and other observables because the integration over momentum would lead to infrared divergences. Therefore, we rewrite the 
decompositions in terms of inverse powers of the particle's energy $\epsilon_{p}=\sqrt{p^{2}+m^{2}}$. Actually, we substitute
$p^{-n}=\epsilon_{p}^{-n}\left[1+\frac{n}{2}\frac{m^{2}}{\epsilon_{p}^{2}}+\mathcal{O}(\epsilon_{p}^{-4})\right]$ and combine the terms with the same power of $\epsilon_{p}$. Obviously, this procedure does not change the 
first two leading order coefficients in each Laurent series while the third one is modified in an appropriate way. Finally, we get Eqs.~(\ref{f-decomp-eps})--(\ref{h-decomp-eps}).

\section{Dimensional regularization}
\label{app-dimensional}


In this appendix, we apply the dimensional regularization in order to extract the divergent part of the corresponding integrals in momentum space. Let us consider the following general integral:
\begin{equation}
I_{\alpha,\beta}=\int\frac{d^{3}\mathbf{p}}{(2\pi)^{3}} (p^{2})^{\alpha} (p^{2}+m^{2})^{\beta}.
\end{equation}
Generalizing it to a $d$-dimensional spacetime [one time dimension and $(d-1)$ space dimensions], we have
\begin{equation}
I^{(d)}_{\alpha,\beta}=\frac{\mu_{r}^{4-d}}{(2\pi)^{d-1}}\int d\Omega_{d-1}\int_{0}^{\infty}dp\,p^{d-2+2\alpha} (p^{2}+m^{2})^{\beta},
\end{equation}
where $\mu_{r}$ is a free parameter with the dimension of mass introduced in order to restore the correct dimension of $I_{\alpha,\beta}$ and $d\Omega_{d-1}$ is the element of the solid angle in $(d-1)$-dimensional space. 
As is well known, the full solid angle equals $\Omega_{d-1}=2\pi^{(d-1)/2}/\Gamma[(d-1)/2]$, where $\Gamma(z)$ is the Euler's gamma function. Performing the change of variables $p^{2}+m^{2}=m^{2}/ \tau$, we obtain
\begin{eqnarray}
I_{\alpha,\beta}&=&\left(\frac{m^{2}}{4\pi\mu_{r}^{2}}\right)^{\frac{d-4}{2}}\frac{m^{2\alpha+2\beta+3}}{8\pi^{3/2}\Gamma\left(\frac{d-1}{2}\right)}\int_{0}^{1}d\tau\,(1-\tau)^{\frac{d-1}{2}+\alpha-1} \tau^{-\frac{d-1}{2}-\alpha-\beta-1}=\nonumber\\
&=&\left(\frac{m^{2}}{4\pi\mu_{r}^{2}}\right)^{\frac{d-4}{2}}\frac{m^{2\alpha+2\beta+3}}{8\pi^{3/2}}\frac{\Gamma\left(\frac{d-1}{2}+\alpha\right)\Gamma\left(-\frac{d-1}{2}-\alpha-\beta\right)}{\Gamma\left(\frac{d-1}{2}\right)\Gamma\left(-\beta\right)},
\end{eqnarray}
where we used the integral representation for the Euler's beta function
\begin{equation}
B(z,w)=\int_{0}^{1}d\tau\,\tau^{z-1}(1-\tau)^{w-1}
\end{equation}
and its relation to the Euler's gamma function
\begin{equation}
B(z,w)=\frac{\Gamma(z)\Gamma(w)}{\Gamma(z+w)}.
\end{equation}
In the subsequent expansion, we will use the well-known recurrence relation of the gamma function
\begin{equation}
\Gamma(z+1)=z\Gamma(z)
\end{equation}
and the asymptotic behavior in the vicinity of its simple pole at $z=0$
\begin{equation}
\Gamma(z)=\frac{1}{z}-\gamma_{E}+\mathcal{O}(z),\qquad\text{for}\ \ z\to 0,
\end{equation}
where $\gamma_{E}\approx 0.577\ldots$ is the Euler-Mascheroni constant.

The main idea of the dimensional regularization is to calculate the integral for an arbitrary $d$ and then consider the limit $d=4-\varepsilon$, $\varepsilon\to 0$, i.e., restore the real dimensionality of the
spacetime. If the initial integral is UV divergent, there will be some terms which blow up at $\varepsilon\to 0$. This divergent part can be easily extracted in the final expression. Let us consider a few 
particular cases:
\begin{equation}
\int\frac{d^{3}\mathbf{p}}{(2\pi)^{3}}\epsilon_{p}\to I^{(d)}_{0,\frac{1}{2}}= -\frac{m^{4}}{16\pi^{2}} \left(\frac{m^{2}}{4\pi\mu_{r}^{2}}\right)^{-\frac{\varepsilon}{2}}\frac{\Gamma\left(\frac{\varepsilon}{2}\right)}{\left(1-\frac{\varepsilon}{2}\right)\left(2-\frac{\varepsilon}{2}\right)}
=-\frac{m^{4}}{32\pi^{2}}\left[\frac{2}{\varepsilon}-\gamma_{E}-\ln\left(\frac{m^{2}}{4\pi\mu_{r}^{2}}\right)+\frac{3}{2}+\mathcal{O}(\varepsilon)\right],
\end{equation}
\begin{equation}
\int\frac{d^{3}\mathbf{p}}{(2\pi)^{3}}\frac{1}{\epsilon_{p}}\to I^{(d)}_{0,-\frac{1}{2}}= -\frac{m^{2}}{8\pi^{2}} \left(\frac{m^{2}}{4\pi\mu_{r}^{2}}\right)^{-\frac{\varepsilon}{2}}\frac{\Gamma\left(\frac{\varepsilon}{2}\right)}{\left(1-\frac{\varepsilon}{2}\right)}
=-\frac{m^{2}}{8\pi^{2}}\left[\frac{2}{\varepsilon}-\gamma_{E}-\ln\left(\frac{m^{2}}{4\pi\mu_{r}^{2}}\right)+1+\mathcal{O}(\varepsilon)\right],
\end{equation}
\begin{equation}
\int\frac{d^{3}\mathbf{p}}{(2\pi)^{3}}\frac{1}{\epsilon_{p}^{3}}\to I^{(d)}_{0,-\frac{3}{2}}= \frac{1}{4\pi^{2}} \left(\frac{m^{2}}{4\pi\mu_{r}^{2}}\right)^{-\frac{\varepsilon}{2}}\Gamma\left(\frac{\varepsilon}{2}\right)
=\frac{1}{4\pi^{2}}\left[\frac{2}{\varepsilon}-\gamma_{E}-\ln\left(\frac{m^{2}}{4\pi\mu_{r}^{2}}\right)+\mathcal{O}(\varepsilon)\right],
\end{equation}
\begin{equation}
\label{int-vi-vj}
\int\frac{d^{3}\mathbf{p}}{(2\pi)^{3}}\frac{v_{i}v_{j}}{\epsilon_{p}^{3}}\to\delta_{ij}\frac{1}{(d-1)} I^{(d)}_{1,-\frac{5}{2}}=\delta_{ij}\frac{1}{3} I^{(d)}_{0,-\frac{3}{2}}= \frac{\delta_{ij}}{12\pi^{2}}\left[\frac{2}{\varepsilon}-\gamma_{E}-\ln\left(\frac{m^{2}}{4\pi\mu_{r}^{2}}\right)+\mathcal{O}(\varepsilon)\right],
\end{equation}
where in the last equation we used the fact that 
\begin{equation}
\langle v_{i}v_{j}\rangle_{\Omega}=\frac{p^{2}}{\epsilon_{p}^{2}}\frac{\delta_{ij}}{d-1}.
\end{equation}

There are many ways to extract the divergent contribution. In particular, in the minimal subtraction (MS) scheme, one subtracts only the term $2/\varepsilon$ which blows up for $\varepsilon\to 0$. However, one may also 
subtract any constant together with this infinite term. We will use the modified minimal subtraction scheme ($\overline{\rm MS}$ scheme), where one subtracts the quantity
\begin{equation}
\label{Upsilon}
\Delta_{\varepsilon}=\frac{2}{\varepsilon}-\gamma_{E}+\ln(4\pi).
\end{equation}

\section{The calculation of energy-momentum tensor}
\label{app-EMT}

The energy-momentum tensor of charged scalar particles produced due to the Schwinger effect has the form
\begin{eqnarray}
\label{T-mu-nu-chi}
T^{\mu\nu}_{\chi}&=&\left<(\mathcal{D}^{\mu}\chi)^{\dagger}(\mathcal{D}^{\nu}\chi)+(\mathcal{D}^{\nu}\chi)^{\dagger}(\mathcal{D}^{\mu}\chi)\right>-g^{\mu\nu}\left<(\mathcal{D}_{\lambda}\chi)^{\dagger}(\mathcal{D}^{\lambda}\chi)-m^{2}|\chi|^{2}\right>+\nonumber\\
&+&2\xi \left<|\chi|^{2}\right>\left(R^{\mu\nu}-\frac{1}{2}R g^{\mu\nu}\right)-2\xi\left<\left(\nabla^{\mu}\nabla^{\nu}-g^{\mu\nu}\nabla_{\lambda}\nabla^{\lambda}\right)|\chi|^{2}\right>,
\end{eqnarray}
where the identity $\delta g^{\alpha\beta}/\delta g_{\mu\nu}=-g^{\mu\alpha}g^{\nu\beta}$ was used. Taking the $00$ component as well as the trace of this energy-momentum tensor, we arrive at Eqs.~(\ref{rho-chi-1})--(\ref{trace-T-chi}). Using Eq.~(\ref{decomposition}) and calculating the vacuum expectation value, we obtain the following results in terms of the mode function $\chi_{\mathbf{k}}$:
\begin{equation}
\rho_{\chi}=\int\frac{d^{3}\mathbf{k}}{(2\pi a)^{3}}\left\{\Big|\dot{\chi}_{\mathbf{k}}-\frac{3}{2}H\chi_{\mathbf{k}}\Big|^{2}+\left[m^{2}-12H^{2}\xi+\frac{(\mathbf{k}-e\mathbf{A})^{2}}{a^{2}}\right]\left|\chi_{\mathbf{k}}\right|^{2}+6H\xi \partial_{0}\left|\chi_{\mathbf{k}}\right|^{2} \right\}.
\end{equation}
\begin{equation}
T_{\chi}=\int\frac{d^{3}\mathbf{k}}{(2\pi a)^{3}}\bigg\{12\Big(\xi-\frac{1}{6}\Big)\Big[\big(|\dot{\chi}_{\mathbf{k}}|^{2}-\Omega_{\mathbf{k}}^{2}|\chi_{\mathbf{k}}|^{2}\big)-3H\Re e\,\big(\chi_{\mathbf{k}}\dot{\chi}_{\mathbf{k}}^{*}\big)\Big]
+\Big[2m^{2}-18\Big(\xi-\frac{1}{6}\Big)\dot{H}\Big]|\chi_{\mathbf{k}}|^{2} \bigg\}.\label{trace-T}
\end{equation}
Finally, using Eq.~(\ref{Bogolyubov-decomposition}), we express the energy density in terms of the kinetic functions and obtain Eqs.~(\ref{rho-chi})--(\ref{T-chi}).

Using the few first terms in the power expansion of the kinetic functions $\mathcal{F}$, $\mathcal{G}$, and $\mathcal{H}$ given by Eqs.~(\ref{f-decomp-eps})--(\ref{h-decomp-eps}), we represent the integrands in 
Eqs.~(\ref{rho-chi})--(\ref{T-chi}) in the following form:
\begin{multline}
\rho_{\chi}=\int\frac{d^{3}\mathbf{p}}{(2\pi)^{3}}\Big\{\epsilon_{p}+\frac{H^{2}}{2\epsilon_{p}}(1-6\xi)+\frac{H(\mathbf{v}\cdot\boldsymbol{\mathcal{E}})}{2\epsilon_{p}^{2}}(1-6\xi)+\\
+\frac{(\mathbf{v}\cdot\boldsymbol{\mathcal{E}})^{2}}{8\epsilon_{p}^{3}}+\frac{H^{2}m^{2}}{2\epsilon_{p}^{3}}(1-6\xi)-\frac{2H\ddot{H}-\dot{H}^{2}+6H^{2}\dot{H}}{8\epsilon_{p}^{3}}(1-6\xi)^{2}+\mathcal{O}(\epsilon_{p}^{-4})\Big\}.
\label{energy-density-divergence}
\end{multline}
\begin{multline}
T_{\chi}=\int\frac{d^{3}\mathbf{p}}{(2\pi)^{3}}\bigg\{\frac{m^{2}}{\epsilon_{p}}+\frac{\dot{H}+H^{2}}{\epsilon_{p}}(1-6\xi)+\frac{\mathbf{v}\cdot\dot{\boldsymbol{\mathcal{E}}}}{2\epsilon_{p}^{2}}(1-6\xi)+\frac{\boldsymbol{\mathcal{E}}^{2}-3(\mathbf{v}\cdot\boldsymbol{\mathcal{E}})^{2}}{2\epsilon_{p}^{3}}+\\
+\frac{m^{2}(2\dot{H}+3H^{2})}{2\epsilon_{p}^{3}}(1-6\xi)-\frac{\dddot{H}+4\dot{H}^{2}+7H\ddot{H}+12H^{2}\dot{H}}{4\epsilon_{p}^{3}}(1-6\xi)^{2} +\mathcal{O}(\epsilon_{p}^{-4})\bigg\}.
\label{trace-divergence}
\end{multline}

The integrals are UV divergent and we apply the dimensional regularization (see Appendix~\ref{app-dimensional}) in order to extract divergent terms. They are given by Eqs.~(\ref{rho-div})--(\ref{T-div}), while the regular 
parts are given by Eqs.~(\ref{rho-chi-reg-1})--(\ref{T-chi-reg-1}).

The effective energy-momentum tensor corresponding to the counterterm (\ref{counterterm-Z3}) reads as
\begin{equation}
\label{EMT-Z3}
T^{\mu\nu}_{Z_{3}}=(Z_{3}-1)\left(F^{\mu\alpha}g_{\alpha\beta}F^{\beta\nu}+\frac{1}{4}g^{\mu\nu}F^{\alpha\beta}F_{\alpha\beta}\right).
\end{equation}
Its 00 component gives the energy density (\ref{rho-Z3}) while the trace $g_{\mu\nu}T^{\mu\nu}_{Z_{3}}$ vanishes.

\end{document}